\documentclass[aip, amsmath, amssymb, reprint,groupedaddress]{revtex4-1}
\usepackage{dcolumn, bm}
\usepackage{graphicx, tikz, orcidlink}
\usepackage{epstopdf}
\usepackage{ascmac}
\usepackage[normalem]{ulem}
\usepackage{cancel}
\usepackage{here}
\usepackage{comment}
\usepackage{mathrsfs, cancel,multirow}

\usepackage[]{hyperref}
\usepackage{xcolor}
\definecolor{AIPBlue}{RGB}{61, 180, 229}
\hypersetup{
colorlinks,
linkcolor=AIPBlue,
citecolor=AIPBlue,
urlcolor=AIPBlue
}

\newcounter{aqctr}
\newenvironment{author-query}
{\refstepcounter{aqctr}\par\vspace{\baselineskip}\noindent
\color{blue}\textbf{Author Query/Comment AQ \arabic{aqctr}.}}
{\par\vspace{\baselineskip}\normalcolor}

\begin{document}
\title{Classical and quantum thermodynamics in a non-equilibrium regime: Application to thermostatic Stirling engine}
\date{Last updated: \today}

\author{Shoki Koyanagi \orcidlink{0000-0002-8607-1699}}\email[Authors to whom correspondence should be addressed: ]{koyanagi.syoki.36z@st.kyoto-u.jp and tanimura.yoshitaka.5w@kyoto-u.jp}\affiliation{Department of Chemistry, Graduate School of Science,
Kyoto University, Kyoto 606-8502, Japan}

\author{Yoshitaka Tanimura \orcidlink{0000-0002-7913-054X}}
\email[Authors to whom correspondence should be addressed: ]{koyanagi.syoki.36z@st.kyoto-u.jp and tanimura.yoshitaka.5w@kyoto-u.jp}
\affiliation{Department of Chemistry, Graduate School of Science,
Kyoto University, Kyoto 606-8502, Japan}

\begin{abstract}
We have developed a thermodynamic theory in the non-equilibrium regime,  which we describe as a thermodynamic system-bath model [S. Koyanagi and Y. Tanimura, J. Chem. Phys. \textbf{160}, 234112 (2024)].  Based on the dimensionless (DL) minimum work principle, non-equilibrium thermodynamic potentials are expressed in terms of non-equilibrium extensive and intensive variables in time derivative form. This is made possible by incorporating the entropy production rate into the definitions of non-equilibrium thermodynamic potentials.  These potentials can be evaluated from the DL non-equilibrium-to-equilibrium minimum work principle, which is derived from the principle of DL minimum work and is equivalent to the second law of thermodynamics. We thus obtain the non-equilibrium Massieu--Planck potentials as entropic potentials and the non-equilibrium Helmholtz--Gibbs potentials as free energies. Unlike fluctuation theorem and stochastic thermodynamics theory, this theory does not require the assumption of a factorized initial condition and is valid in the full quantum regime where the system and bath are quantum mechanically entangled.
Our results are numerically verified by simulating a thermostatic Stirling engine consisting of two isothermal processes and two thermostatic processes using the quantum hierarchical Fokker--Planck equations and the classical Kramers equation derived from the thermodynamic system-bath model. We then show that, from weak to strong system-bath interactions, the thermodynamic process can be analyzed using a non-equilibrium work diagram analogous to the equilibrium one for given time-dependent intensive variables. The results can be used to develop efficient heat machines in non-equilibrium regimes. 
\end{abstract}
\maketitle

\section{Introduction}
\label{sec.intro}
Ever since Carnot explored the efficiency of heat engines 200 years ago,\cite{Carnot1824} there have been longstanding attempts to study thermodynamics in non-equilibrium regimes driven by academic curiosity and practical interest. 
In particular, recent advances in nanotechnology have led to increased interest in the study of the thermodynamics of microscopic systems.
\cite{PhysRevLett.97.180402,Kosloff2014AnnRev,GalperinPhysRevB2015,PhysRevB.91.224303,PhysRevX.5.031044,WhitneyPhysRevB2018,NoriOtto2007,NoriDemon2009,OttoNori2020,NoriDemon2009,PhysRevBUncertenty2020,PRXQuantum.2.030202,RevModPhys.81.1665,RevModPhys.83.771,sagawa20232law,Schmiedl_2008,Stochastic_Seifert_2012,Esposito2010,Esposito2011,PhysRevA.106.062209,strasberg2022quantum,RaoEsposit2018,PhysRevE.105.064124,HanggiRevModPhys2020,Kosloff_2023,usui2024microscopic,PhysRevA.106.062209,PhysRevResearch.5.043274,ST20JCP,ST21JPSJ,KT15JCP,KT16JCP} 

As a kinetic theory, such phenomena are typically explained using open quantum dynamics theories based on the system--bath (SB) model.\cite{SpinBosonLettett,CALDEIRA1983587,GRABERT1988115,Weiss2012}  Although thermodynamics is a system-independent theory, it is derived from the presence of a heat bath and is consistent with theories based on the SB model.\cite{HanggiRevModPhys2020} The main (subsystem) system of the SB model can be described from a simple two-level system to complex molecular systems with many degrees of freedom, and it is also possible to take the classical limit in the case of a system described in phase space.\cite{CALDEIRA1983587,GRABERT1988115,Weiss2012,TK89JPSJ1,T90PRA,T14JCP,T06JPSJ,T20JCP}  Under the condition of a heat bath with infinite degrees of freedom, the total system irreversibly relaxes toward its equilibrium state in time, so the second law of thermodynamics is naturally obeyed without the need for the assumption of ergodicity as a dynamical system.\cite{T06JPSJ,T20JCP,T14JCP,T15JCP}

However, to investigate whether the thermodynamic laws are satisfied even in quantum cases where the subsystem and bath are quantum mechanically entangled, it is necessary to treat the bath in a non-Markovian, non-perturbative, and non-factorized manner so that the thermal equilibrium state of the total system satisfies the energy conservation law, including the SB interaction. Thus, equations of motion derived using the Markovian or rotational wave (or secular) approximation, such as the Lindblad equation and quantum master equation, can only be applied to high-temperature regions where the subsystem exhibits semiclassical dynamics.\cite{T06JPSJ,T20JCP,T14JCP,T15JCP}  
Therefore, the validity of theories based on the Markov assumption\cite{Keiji2022Markov} or those based on factorized initial conditions, such as the fluctuation theorem\cite{RevModPhys.81.1665,RevModPhys.83.771,sagawa20232law} and stochastic thermodynamics\cite{Schmiedl_2008,Stochastic_Seifert_2012,Esposito2010,Esposito2011,strasberg2022quantum,RaoEsposit2018,PhysRevE.105.064124,PhysRevA.106.062209} in the quantum case, should be carefully examined.

Over the past 30 years, several methodologies have been developed to accurately describe the effects of quantum entanglement in condensed systems while maintaining strict energy conservation to satisfy the first law of thermodynamics.  Such methods   include the hierarchical equations of motion (HEOM), \cite{TK89JPSJ1,T90PRA,IT05JPSJ,T06JPSJ,T20JCP,T14JCP,T15JCP}
the quasi-adiabatic path integral (QUAPI),\cite{Makri95, Makri96, Makri96B,Thorwart00,JadhaoMakri08,MakriJCP2014,Reichman2010} and  multiconfigurational time-dependent Hartree (MCTDH),\cite{ML-MCTDH1,ML-MCTDH2, WangTHoss2} in historical order.
Among these, the HEOM approach is ideal for thermodynamic investigations, not only because it can perform numerically ``exact'' dynamic simulations but also because it can evaluate the change in energy of the heat bath and the SB interaction separately, even for processes far from equilibrium.\cite{KT15JCP,KT16JCP,ST20JCP,ST21JPSJ,PhysRevB.95.064308,YiJing2022,Petruccione2023,Strunz2024,KT22JCP1,KT22JCP2,KT24JCP4,KT24JCP1,KT24JCP3}
Note that although not numerically ``exact'', several non-perturbative approaches have also been developed specifically for thermodynamic systems.\cite{PhysRevB.97.205405,PhysRevE.101.050101,PhysRevB.109.085408}

While the difficulty of numerical simulation has been solved, a fundamental difference exists between open quantum dynamics theory, which is based on a first-principles description of kinetic systems from a microscopic perspective, and thermodynamic theory, which is based on a phenomenological description of thermal systems from a macroscopic perspective. For example, in quantum mechanics, observables are defined as expectation values, whereas in thermodynamics, they are described by macroscopic intensive and extensive variables. Furthermore, quantum mechanics is a formalism for time evolution, whereas thermodynamics deals primarily with static and quasi-static states near thermal equilibrium. 
In this respect, most quantum thermodynamics studies are merely kinetic simulations of open quantum dynamics, and their relation to thermodynamics has not been studied in depth. 
Thus, the fundamental difference between microscopic quantum mechanics and macroscopic thermodynamics raises many open questions, such as whether the Carnot limit can be violated in a quantum case or the existence of Maxwell's demon. 

The virtue of thermodynamics lies in its ability to describe macroscopic thermal phenomena resulting from complex microscopic interactions in a system-independent manner, as changes in thermodynamic potentials described as interrelated intensive and extensive variables through Legendre transformations. This virtue should be preserved when developing quantum thermodynamic theory rather than open quantum dynamical theory, although in either case, the theory must be specific to the SB model.

The assumption of a factorized initial state is essential for the application of stochastic thermodynamics and the fluctuation theorem, so that these theories cannot describe the transitions between states in which the system and the bath are entangled.  Also, in practice 
many quantum thermodynamics arguments treat a heat bath perturbatively, assuming Markovian time evolution,\cite{RevModPhys.81.1665,RevModPhys.83.771,sagawa20232law,Schmiedl_2008,Stochastic_Seifert_2012} it has been found that the Gibbs energy can be obtained directly from kinetic simulations even in the case of non-Markovian and non-perturbative SB interaction at low temperature, where quantum entanglement between the system and the bath plays an essential role.\cite{ST20JCP,ST21JPSJ,KT22JCP1,KT22JCP2} This approach based on the minimum work principle (or Kelvin--Planck statement), expressed as $W(t) \ge \Delta G(t)$, where $W ( t )$ is the work done by the outside on the subsystem by external fields and $\Delta G(t)$ is the change in free energy, by evaluating the work in a quasi-static process,\cite{ST20JCP,ST21JPSJ} which allows us to draw a work diagram corresponding to the $P$--$V$ diagram.\cite{KT22JCP1,KT22JCP2} 
However, even with this approach, the contributions of temperature $T$ and entropy $S$ to the thermodynamic potential expressed as $TdS$ and $SdT$ cannot be evaluated because the minimum work principle is defined as an isothermal process $dT=0$.

To overcome this limitation, we developed a thermostatic SB model that is defined by a system coupled to multiple heat baths at different temperatures.\cite{KT24JCP1,KT24JCP3} 
We then extended the minimum work principle to thermostatic processes in a dimensionless (DL) form (the DL minimum work principle) as ${\tilde W}(t) \ge \Delta \Xi (t)$, where ${\tilde W}(t)  \equiv \beta(t) W(t) $ is the DL (entropic) work,  $\beta(t)\equiv 1/k_{\rm B} T(t)$ (where $k_{\rm B}$ is the Boltzmann constant) is the time-dependent inverse temperature, and $\Delta \Xi$ is the change in the DL Planck potential.\cite{KT24JCP1} 
 Not only intensive variables but also extensive variables, which are related by time-dependent Legendre transformations, were introduced as quantum expectation values of the SB system. 
 
The validity of these results was verified using the numerically ``exact'' HEOM formalism. For thermodynamic studies, the HEOM approach has been used for spin-boson-based systems, taking advantage of the ability to evaluate the energy changes of the system, interaction, and bath, respectively, even under non-perturbative and non-Markovian conditions.\cite{KT15JCP,KT16JCP,ST20JCP,ST21JPSJ,KT22JCP1,KT22JCP2,KT24JCP4}  Here, we employed the HEOM formalism for an anharmonic quantum Brownian model to construct a thermodynamic theory that is valid for both classical and quantum cases.  Restricting to the case of an Ohmic spectral distribution function (SDF), we derived the thermostatic quantum Fokker--Planck equations (T-QFPE)\cite{KT24JCP1,KT24JCP3} on the basis of the low-temperature quantum Fokker--Planck equations (LT-QFPE) in the quantum case and the Kramers equation in the classical case.\cite{IT19JCTC}  In the classical and high-temperature limits, the T-QFPE are equivalent to the Langevin equation, where a Markovian description can be applicable, but at low temperatures, owing to quantum entanglement with the bath (bathentanglement),\cite{T20JCP} the subsystem follows non-factorial and non-Markovian dynamics, and its equilibrium state deviates from the Boltzmann distribution. This indicates that thermodynamics in the fully quantum regime cannot be described by a theory based on the Markovian assumption.\cite{KT24JCP3,KT24JCP4}

Although these results were restricted to quasi-static cases, the  thermodynamic potentials, intensive and extensive variables, and Legendre transformations were defined in such a way that they hold for any non-equilibrium process. 
Taking advantage of this, we extend our thermodynamic theory here to the non-equilibrium regime.

The remainder of this paper is organized as follows. 
In Sec.~\ref{sec:Thermodynamics}, our previous results  about thermodynamics as applied to work in a system-independent manner are summarized. We then derive the principle of non-equilibrium DL minimum work to obtain the DL Massieu--Planck potential and Helmholtz--Gibbs potentials in the non-equilibrium regime.   
Results are verified in Sec.~\ref{sec:Simulation} by numerical simulations using the thermostatic SB model. Finally, Sec.~\ref{sec:conclude} presents  concluding remarks.

\section{Reflections on motive power of heat}
\label{sec:Thermodynamics}

In our previous paper,\cite{KT24JCP1} we presented system-specific thermodynamic laws described as a system-bath model on the basis of open quantum dynamics theory. Here, we develop the same laws by introducing several thermodynamic ``statements'' without going into the details of the system, as in traditional thermodynamic theory.  By doing so, we clarify the distinctive features of thermodynamic theory that allow it to treat systems in non-equilibrium regimes.

\subsection{Laws of thermodynamics applied to work}
\label{sec:DLWHI}

We consider a thermodynamic system consisting of subsystem A and heat bath B at the inverse temperature $\beta(t)$.  The presence of a heat bath temperature can be regarded as a consequence of {\bf the zeroth law of thermodynamics}, which states the existence of a unique equilibrium state.\cite{Oono2017}
Although not well known, {\bf the minus first law of thermodynamics} was introduced some time ago to state the existence of time-irreversible processes toward a unique equilibrium state.\cite{brown2001minus1st,uffink2001minus1st2} The dynamics of the SB model comply with this law.

An important statement of thermodynamics is that thermodynamic systems are described in terms of intensive and extensive variables. In particular, extensive variables that are proportional to the size are essential. This statement is called the fourth law.\cite{Oono2017} However, this is the premise of thermodynamics, and we would like to call it {\bf  the minus second law of thermodynamics} because this is the central dogma in the classical and quantum thermodynamics that we are constructing.

The energy of the subsystem corresponds to the internal energy and is expressed as $U_{\rm A} ( t )$, which changes with time owing to changes in the bath temperature. Internal energy is an extensive variable, whereas  inverse temperature is an intensive variable. The external perturbation considered here is expressed as $-x(t) X_{\rm A}(t)$, where $x(t)$ and $X_{\rm A}(t)$ are  intensive and extensive variables, respectively. In traditional thermodynamics, $x(t)$ is derived from the Euler relation. We then introduce the total energy, which is related to the enthalpy by ${H}_{\rm A} ( t ) =  U_{\rm A} ( t ) - x ( t ) X_{\rm A} ( t )$, which can also be regarded as the Legendre transformation between internal energy and enthalpy. 

Because $U_{\rm A} ( t )$ is the conjugate variable of $\beta(t)$, we introduce the DL (or entropic) internal energy denoted by $\tilde{U}_{\rm A} ( t )$, where $\tilde{B}(t)\equiv \beta(t)B(t)$ for any variable $B(t)$. Because $\beta(t)$ diverges for $T \rightarrow 0$, states with $T=0$ do not exist. Thus the introduction of DL variables corresponds to {\bf the third law of thermodynamics}.

From the time derivative of the DL internal energy $\tilde{U}_{\rm A} ( t )$, we obtain
\begin{eqnarray}
\label{eq:DLFirstLawU}
\frac{d \tilde{U}_{\rm A} ( t )}{d t}
= \frac{d \tilde{W}^{ext}_{\rm A} ( t )}{d t} + \frac{d \tilde{Q}_{\rm A}^{ext} ( t )}{d t} ,
\end{eqnarray}
where 
\begin{eqnarray}
\label{eq:DefWext0}
\frac{d \tilde{W}^{ext}_{\rm A} ( t )}{d t} = U_{\rm A} ( t ) \frac{d \beta ( t )}{d t} + \tilde{x} ( t ) \frac{d  X_{\rm A} ( t )}{d t} 
\end{eqnarray}
and
\begin{eqnarray}
\label{eq:DefDLQext}
\frac{d \tilde{Q}_{\rm A}^{ext}  ( t )}{d t} \equiv \beta ( t ) \frac{d U_{\rm A} ( t )}{d t} 
- \tilde{x}( t ) \frac{d X_{\rm A} ( t )}{d t},
\end{eqnarray}
and $\tilde{W}^{ext}_{\rm A} ( t )$ and $ \tilde{Q}_{\rm A}^{ext}  ( t )$ represent the DL (or entropic) extensive work defined as being done by the outside and DL (or entropic) extensive heat, respectively. The time derivative of $\tilde{W}^{ext}_{\rm A} ( t )$ corresponds to the DL extensive power. For the enthalpy, we have
\begin{eqnarray}
\label{eq:DLFirstLawH}
\frac{d \tilde{H}_{\rm A} ( t )}{d t}=\frac{d \tilde{W}_{\rm A}^{int} ( t )}{d t}+\frac{d \tilde{Q}_{\rm A}^{ext}  ( t )}{d t},
\end{eqnarray}
where
\begin{eqnarray}
\label{eq:DefWint}
\frac{d \tilde{W}_{\rm A}^{int} ( t )}{d t} \equiv U_{\rm A} ( t ) \frac{d \beta ( t )}{d t}
- X_{\rm A} ( t ) \frac{d \tilde{x} ( t )}{d t}
\end{eqnarray}
and $\tilde{W}^{int}_{\rm A} ( t )$ represents the DL intensive work. The time derivative of $\tilde{W}^{int}_{\rm A} ( t )$ then corresponds to the DL intensive power.  Equations~\eqref{eq:DLFirstLawU} and~\eqref{eq:DLFirstLawH} correspond to {\bf the first law of thermodynamics}.  

From the definitions, these intensive and extensive variables satisfy the time-dependent Legendre (TDL) transformation expressed as follows:
\begin{eqnarray}
\label{eq:DefWext}
\frac{d \tilde{W}_{\rm A}^{int} ( t )}{d t} = \frac{d \tilde{W}^{ext}_{\rm A} ( t )}{d t} - \frac{d}{d t} \left[ \tilde{x} ( t ) X_{\rm A} ( t ) \right].
\end{eqnarray}
Equation~\eqref{eq:DLFirstLawU} can also be expressed in the form of a TDL transformation as
\begin{eqnarray}
\label{eq:DefWQ}
\frac{d \tilde{W}_{\rm A}^{ext} ( t )}{d t} = - \frac{d \tilde{Q}^{ext}_{\rm A} ( t )}{d t} 
+ \frac{d}{d t} \left[ \beta ( t ) U_{\rm A} ( t ) \right] .
\end{eqnarray}
The DL intensive heat $\tilde{Q}_{\rm A}^{int}  ( t )$ can also be defined from $\tilde{Q}_{\rm A}^{ext}  ( t )$ using the TDL [Eq. \eqref{eq:DefWQintLe} in Appendix~\ref{sec:QTP}]. The above equations, including the first law of thermodynamics, are described in terms of intensive variables $\beta ( t )$ and $\tilde{x} ( t )$ and extensive variables $U_{\rm A} ( t )$ and $X_{\rm A} ( t )$ and hold for any non-equilibrium processes. While these intensive and extensive variables are the state variables because they are kinetic observables at time $t$, work and heat are not state variables.

As {\bf the second law of thermodynamics}, we adopt {\bf the DL minimum work principle} for the subsystem moving from one equilibrium state to another, thereby extending the Kelvin--Planck statement for isothermal processes to thermostatic processes.
For work defined as being performed from the outside on the subsystem, we have
\begin{eqnarray}
\label{eq:MinimumExtWork0}
\tilde{W}_{\rm A}^{ext} \geq -\Delta \Phi_{\rm A}^{\rm qst},
\end{eqnarray}
where $\Phi_{\rm A}^{\rm qst}$ is the DL Massieu potential \cite{massieu1869, Callen1991} and equality holds under quasi-static changes in the natural variables $\beta(t)$ and $X_{\rm A}(t)$ as $\Phi_{\rm A}^{\rm qst}[\beta^{\rm qst},  X_{\rm A}^{\rm qst}]=\tilde{W}_{\rm A}^{ext} [\beta^{\rm qst},  X_{\rm A}^{\rm qst}]$.  The above inequality states that for any process occurring between two states, the work done by the outside is minimized if the process is quasi-static (or reversible) because no energy is dissipated in the heat bath. Note that if we define work as being done by the subsystem on the outside, the inequality in Eq.\eqref{eq:MinimumExtWork0} is reversed, and the relationship is called the maximum work principle (on the outside). Although not mathematically rigorous, the derivation of Eq. \eqref{eq:MinimumExtWork0} is described in Appendix A of Ref. \onlinecite{KT24JCP1}. Using the TDL transformation~\eqref{eq:DefWext}, we also obtain the Planck potential \cite{Planck1922, Callen1991}  in terms of natural variables expressed as $\Xi_{\rm A}^{\rm qst} [\beta^{\rm qst},  \tilde{x}^{\rm qst}]$ (see Appendix~\ref{sec:QTP}).
Similar expressions for Planck functions were derived on the basis of stochastic thermodynamics.\cite{Esposito2011,RaoEsposit2018, strasberg2022quantum} However, our expression was obtained from the well-defined kinetic system and is valid even under non-factorized SB interactions, where bathentanglement plays an essential role.
Thus, the value of $\Delta \Phi_{\rm A}^{\rm qst}$ appearing in Eq. \eqref{eq:MinimumExtWork0} 
is generally different from the value evaluated from stochastic thermodynamics. 

For the DL extensive heat $\tilde{Q}_{\rm A}^{ext}  ( t )$, we have
\begin{eqnarray}
\label{eq:MinimumintQ0}
\tilde{Q}_{\rm A}^{ext} \leq \Delta \Lambda_{\rm A}^{\rm qst},
\end{eqnarray}
which corresponds to {\bf the principle of maximum entropy}, where $\Lambda_{\rm A}^{\rm qst}[U_{\rm A}^{\rm qst}, {X}_{\rm A}^{\rm qst}]$ is the entropic potential. 
Because both natural variables are extensive, this potential is fundamental and we call it the Massieu entropy (M-entropy). Using the TDL transformation given by Eq. \eqref{eq:DefWQintLe}, we also obtain the Planck entropy (P-entropy) defined by $\Gamma_{\rm A}^{\rm qst} [U_{\rm A}^{\rm qst},  \tilde{x}^{\rm qst}]$ from $\tilde{Q}_{\rm A}^{int}  ( t )$ (see Appendix~\ref{sec:QTP}).   Similar to the entropy obtained from the partition function in statistical physics, this entropy is a function of the intensive variable $\tilde{x}^{\rm qst}$.

Here, as the conjugate variable for $\beta(t)$, we chose the internal energy $U_{\rm A}^{\rm qst}$, which is considered a fundamental variable in thermodynamics. As demonstrated in our previous paper,\cite{KT24JCP1} enthalpy $H_{\rm A}^{\rm qst}$ could also be chosen as the conjugate variable for $\beta(t)$. In such a case, there are also two entropies, one involving both extensive variables (Clausius entropy or C-entropy) and one intensive variable with respect to external forces (Boltzmann entropy or B-entropy).  The B-entropy and M-entropy values are related to the others by Legendre transformations between $U_{\rm A}^{\rm qst}$ and $H_{\rm A}^{\rm qst}$ and the values do not change, although the former includes one intensive variable, whereas the latter includes extensive variables only. However, there is no explicit relationship between C-entropy and M-entropy, because the Legendre transformations between two entropies in the $U_{\rm A}^{\rm qst}$ representation and the $H_{\rm A}^{\rm qst}$ representation are different. (See Appendix D in Ref. \onlinecite{KT24JCP1}). The Massieu and Planck potentials are
equivalent to the entropy potentials introduced earlier.\cite{Guggenheim1986} 
 
The total differential form of the Helmholtz--Gibbs potentials can be derived from the Massieu--Planck potentials using the definitions $F^{\rm qst}_{\rm A} ( t ) = - \Phi_{\rm A}^{\rm qst} ( t ) / \beta^{\rm qst}(t)$ and $G^{\rm qst}_{\rm A} ( t ) = - \Xi_{\rm A}^{\rm qst} ( t ) / \beta ^{\rm qst}(t)$.  The results are summarized in Table~\ref{table:Potential} in Appendix~\ref{sec:QTP}.

\subsection{Thermodynamic potentials in a non-equilibrium regime}
\label{sec:NeqDLPotentials}

\subsubsection{The DL non-equilibrium minimum work principle}
\label{sec:NE2EP}

On the basis of Eq.~\eqref{eq:MinimumExtWork0}, we define thermodynamic potentials applicable to the non-equilibrium regime.  Consider a non-equilibrium state $A$ and equilibrium state $n$. {\bf The DL non-equilibrium-to-equilibrium minimum work principle} for $A \rightarrow n$ is expressed as (see Appendix~\ref{sec:DNEMWP})
\begin{eqnarray}
\label{eq:MinimumPrinciple}
( \tilde{W}_{\rm A}^{ext} )_{{A} \rightarrow n} \geq - ( \Delta \Phi_{A}^{\rm neq} )_{A \rightarrow n},
\end{eqnarray}
where $\Phi_{\rm A}^{\rm neq} ( t )$ is the non-equilibrium Massieu potential. 
This inequality indicates that the path from non-equilibrium to equilibrium has a lower bound on work. In engineering, the effective energy from non-equilibrium to equilibrium is referred to as {\bf exergy}.\cite{Rant1956Exergie} The non-equilibrium thermodynamic potential introduced here can be regarded as a generalization of it. 

The difference between $(\tilde{W}_{\rm A}^{ext} )_{{A} \rightarrow n}$ and $- ( \Delta \Phi_{A}^{\rm neq} )_{{A} \rightarrow n}$ corresponds to the entropy production defined as 
\begin{eqnarray}
\label{eq:DefWasteHeat}
(\Sigma_{\rm A})_{{A} \rightarrow n} = ( \tilde{W}_{\rm A}^{ext} )_{{A} \rightarrow n}
+ ( \Delta \Phi_{\rm A}^{\rm neq} )_{{A} \rightarrow n} \geq 0 .
\end{eqnarray}
For two non-equilibrium states $A$ and $B$, we have
$(\Sigma_{\rm A})_{{A} \rightarrow B}$ = $(\Sigma_{\rm A})_{{A} \rightarrow n} - (\Sigma_{\rm A})_{{B} \rightarrow n}$.  Thus, the entropy production rate for any non-equilibrium process at times $t$ and $t + d t$ (where $d t$ is an infinitesimal time) is expressed as
\begin{eqnarray}
\label{eq:DiffWasteHeat}
\frac{d \Sigma_{\rm A} ( t )}{d t} = \frac{d \tilde{W}_{\rm A}^{ext} ( t )}{d t}
+ \frac{d \Phi_{\rm A}^{\rm neq} ( t )}{d t} .
\end{eqnarray}
As shown in Appendix~\ref{sec:DNEMWP}, the entropy production rate is alway positive, i.e., $d \Sigma_{\rm A} ( t ) / d t \geq 0$. Thus, we have the inequality $( \tilde{W}_{\rm A}^{ext} )_{A \rightarrow B}^{\rm min} \geq  - (\Delta \Phi_{\rm A}^{\rm neq})_{\rm A \rightarrow B}$, or
\begin{eqnarray}
\label{eq:NeqDLExtWorkIneq0-text}
\frac{d \tilde{W}_{\rm A}^{ext} ( t )}{d t} \geq - \frac{d \Phi_{\rm A}^{\rm neq} ( t )}{d t},
\end{eqnarray}
which we call {\bf the DL non-equilibrium minimum work principle}.
While $\tilde{W}_{\rm A}^{ext} ( t )$ and $\Sigma_{\rm A} ( t )$ are not state variables because they depend on a path, $\Phi_{\rm A}^{\rm neq} ( t )$ is a state variable defined by the non-equilibrium-to-equilibrium minimum work path.  When state $B$ is on the non-equilibrium-to-equilibrium minimum work path from $A$ to $n$,  equality in Eq.~\eqref{eq:NeqDLExtWorkIneq0-text} holds.

Using Eqs.~\eqref{eq:DefWext0} and~\eqref{eq:DiffWasteHeat}, we obtain the time derivative form of the non-equilibrium Massieu potential as
\begin{eqnarray}
\label{eq:DiffNeqMassieu}
\frac{d \Phi_{\rm A}^{\rm neq} ( t )}{d t} = - U_{\rm A}^{\rm neq} \frac{d \beta ( t )}{d t}
-\tilde{x} ( t ) \frac{d  {X}_{\rm A}^{\rm neq} ( t )}{d t} + \frac{d\Sigma_{\rm A} ( t )}{d t}.
\end{eqnarray}
The convexity of DL potentials is discussed on the basis of the SB model in Appendix~\ref{sec:Convexity}. Thus, these can be regarded as thermodynamic potentials.

Using Eqs.~\eqref{eq:DefWext}, we have the non-equilibrium Planck potential expressed as
\begin{eqnarray}
\label{eq:DiffNeqPlanck}
\frac{d \Xi_{\rm A}^{\rm neq} ( t )}{d t} = - U_{\rm A}^{\rm neq} \frac{d \beta ( t )}{d t}
+  {X}_{\rm A}^{\rm neq}  ( t ) \frac{d \tilde{x} ( t )}{d t} + \frac{d\Sigma_{\rm A} ( t )}{d t}.
\end{eqnarray}
The M-entropy $\Lambda_{\rm A}^{\rm neq} ( t )$ and P-entropy $\Gamma_{\rm A}^{\rm neq} ( t )$ can be evaluated from the TDL transformations~\eqref{eq:DefWext} and ~\eqref{eq:DefWQ}. 
The non-equilibrium potentials satisfy the following Legendre transformations:
\begin{eqnarray}
\label{eq:neqLegendreP-M}
\Xi_{\rm A}^{\rm neq} ( t ) = \Phi_{\rm A}^{\rm neq} ( t ) + \tilde{x} ( t )  {X}_{\rm A}^{\rm neq} ( t ) 
\end{eqnarray}
and
\begin{eqnarray}
\label{eq:neqLegendreE-M}
\Lambda_{\rm A}^{\rm neq} ( t ) = \Phi_{\rm A}^{\rm neq} ( t ) + \beta ( t ) U_{\rm A}^{\rm neq} ( t ).
\end{eqnarray}

The time derivative forms of the DL non-equilibrium entropic potentials are summarized in Table~\ref{table:NeqDLPotential}.  Note that the entropic potentials, the intensive variables, and the non-equilibrium extensive variables are state variables, whereas their time derivatives are not state variables.

As shown in Appendix~\ref{sec:neq-qstIneq}, the non-equilibrium Massieu potential is always smaller than the quasi-static Massieu potential.  This indicates that the non-equilibrium Massieu potential is minimum when the state is equilibrium.  In the quasi-static case, Table ~\ref{table:NeqDLPotential} is reduced to Table~\ref{table:DLPotential} (Appendix~\ref{sec:QTP}).  Thus, we can regard these potentials as  extensions of the thermodynamic potentials to a non-equilibrium regime.  

From Eqs.~\eqref{eq:DiffWasteHeat} and~\eqref{eq:neqLegendreE-M}, the entropy production rate is expressed using the non-equilibrium M-entropy as
\begin{eqnarray}
\label{eq:WasteHeat2}
\frac{d\Sigma_{\rm A} ( t )}{d t} = \frac{d \Lambda_{\rm A}^{\rm neq} ( t )}{d t} 
- \frac{d \tilde{Q}_{\rm A}^{ext} ( t )}{d t}.
\end{eqnarray}

\begin{table*}[!t]

\caption{\label{table:NeqDLPotential} Time derivative forms of the non-equilibrium (neq)  Massieu--Planck potentials  as  functions of intensive variables $\beta ( t )$ and  $\tilde{x} ( t )$ and extensive variables $U_{\rm A}^{\rm neq} ( t )$ and  $X_{\rm A}^{\rm neq} ( t )$. Of the DL entropies,  the commonly used one, which we call the Massieu entropy (M-entropy), involves only extensive variables and is denoted by $\Lambda_{\rm A}^{\rm neq}[U_{\rm A}^{\rm neq}(t), {X}_{\rm A}^{\rm neq}(t)]$, whereas the less widely used one, which we call the Planck entropy (P-entropy),  is denoted by $\Gamma_{\rm A}^{\rm qst} [ U_{\rm A}^{\rm qst}, \tilde{x} ]$.  Because heat is always lost in non-equilibrium processes, the entropy production rate $d\Sigma_{\rm A} /dt $ appears in the equations. Each potential is related to others via Legendre transformations shown in the final column. }
\begin{ruledtabular}
\begin{tabular}{lcccc}
neq DL Thermodynamic pot. & Differential Form  & Natural var. & Legendre Transformation
\\
 \hline
Massieu & $\frac{d}{d t} \Phi_{\rm A}^{\rm neq} = - U_{\rm A}^{\rm neq}  \frac{d}{d t} \beta - \tilde{x} \frac{d}{d t} {X}_{\rm A}^{\rm neq} + \frac{d}{d t} \Sigma_{\rm A}$  & $\beta , {X}_{\rm A}^{\rm neq}$ & $\cdots$
\\
Planck & $\frac{d}{d t} \Xi^{\rm neq}_{\rm A} = - U_{\rm A}^{\rm neq}  \frac{d}{d t} \beta + {X}_{\rm A} ^{\rm neq} \frac{d}{d t} \tilde{x} + \frac{d}{d t} \Sigma_{\rm A}$ & $\beta, \tilde{x}$ 
& $\Xi_{\rm A}^{\rm neq} = \Phi_{\rm A}^{\rm neq} + \tilde{x} {X}_{\rm A}^{\rm neq}$ 
\\
M-Entropy & $\frac{d}{d t} \Lambda_{\rm A}^{\rm neq} = \beta \frac{d}{d t} U_{\rm A}^{\rm neq}  - \tilde{x}  \frac{d}{d t} X_{\rm A}^{\rm neq} + \frac{d}{d t} \Sigma_{\rm A} $ & $U_{\rm A}^{\rm neq} , X_{\rm A}^{\rm neq}$ & $\Lambda_{\rm A}^{\rm neq} = \Phi_{\rm A}^{\rm neq} + \beta U_{\rm A}^{\rm neq} $\\
P-Entropy & $\frac{d}{d t} \Gamma_{\rm A}^{\rm neq} = \beta \frac{d}{d t} U_{\rm A}^{\rm neq}  + X_{\rm A}^{\rm neq} \frac{d}{d t} \tilde{x} + \frac{d}{d t} \Sigma_{\rm A}$ & $U_{\rm A}^{\rm neq} , \tilde{x}$ & $\Gamma_{\rm A}^{\rm neq} = \Lambda_{\rm A}^{\rm neq} + \tilde{x} {X}_{\rm A}^{\rm neq} $
\end{tabular}
\end{ruledtabular}
\end{table*}

\subsubsection{Non-equilibrium Helmholtz--Gibbs potentials}


\begin{table*}[!t]
\caption{\label{table:NeqPotential} Time derivative forms of the non-equilibrium (neq) Helmholtz--Gibbs potentials as functions of intensive variables $T ( t )$ and  $x ( t )$ and extensive variables $S_{\rm A}^{\rm neq}(t)$ and  $X_{\rm A}^{\rm neq}(t)$, which are interrelated through the Legendre transformations shown in the final column.  Because heat is always lost in non-equilibrium processes, the waste heat $Q_{\rm A}^{\rm wst}$ appears in the equations.}
\begin{ruledtabular}
\begin{tabular}{lccc}
neq Thermodynamic pot. & Differential Form & Natural var. & Legendre Transformation
\\ 
\hline
Helmholtz Energy & 
$\frac{d}{d t} F^{\rm neq}_{\rm A}
= - S^{\rm neq}_{\rm A} \frac{d}{d t} T + x \frac{d}{d t} X_{\rm A} ^{\rm neq} + \frac{d}{d t} Q_{\rm A}^{\rm wst}$
& $T , X_{\rm A}^{\rm neq}$ & -
\\
Gibbs Energy  &
$\frac{d}{d t} G^{\rm neq}_{\rm A} =
- S^{\rm neq}_{\rm A} \frac{d}{d t} T - X_{\rm A} ^{\rm neq} \frac{d}{d t} x + \frac{d}{d t} Q_{\rm A}^{\rm wst}$
& $T ,  x$ & $G^{\rm neq}_{\rm A} = F^{\rm neq}_{\rm A} - x X_{\rm A}^{\rm neq} $
\\
Internal Energy &
$\frac{d}{d t} U^{\rm neq}_{\rm A} = 
T \frac{d}{d t}  S^{\rm neq}_{\rm A} + x \frac{d}{d t} X_{\rm A}^{\rm neq} + \frac{d}{d t} Q_{\rm A}^{\rm wst}$
& $S_{\rm A}^{\rm neq} , X_{\rm A}^{\rm neq}$ & $U^{\rm neq}_{\rm A} = F^{\rm neq}_{\rm A} + T S^{\rm neq}_{\rm A}$
\\
Enthalpy &
$\frac{d}{d t} H^{\rm neq}_{\rm A} = 
T \frac{d}{d t} S^{\rm neq}_{\rm A} - X_{\rm A}^{\rm neq}  \frac{d}{d t} x + \frac{d}{d t} Q_{\rm A}^{\rm wst}$
& $S_{\rm A}^{\rm neq} , x$ 
& $H^{\rm neq}_{\rm A} = U^{\rm neq}_{\rm A} - x X_{\rm A}^{\rm neq} $
\end{tabular}
\end{ruledtabular}
\end{table*}

We introduce non-equilibrium Helmholtz and Gibbs energies defined as $F^{\rm neq}_{\rm A} ( t ) = - \Phi_{\rm A}^{\rm neq} ( t ) / \beta(t)$ and $G^{\rm neq}_{\rm A} ( t ) = - \Xi_{\rm A}^{\rm neq} ( t ) / \beta (t)$. 
Because $d\beta(t)/dt= - ( 1 / k_{\rm B} T^2 ( t ) ) d T ( t ) / d t$, we obtain the time derivative forms of these from Eqs.~\eqref{eq:DiffNeqMassieu} and~\eqref{eq:DiffNeqPlanck} as
\begin{equation}
\frac{d F_{\rm A}^{\rm neq} ( t )}{d t} = - S_{\rm A}^{\rm neq} ( t ) \frac{d T ( t )}{d t}
+ x ( t ) \frac{d X_{\rm A} ^{\rm neq}( t )}{d t} + \frac{d Q_{\rm A}^{\rm wst} ( t )}{d t}
\end{equation}
and
\begin{equation}
\frac{d G_{\rm A}^{\rm neq} ( t )}{d t} = - S_{\rm A}^{\rm neq} ( t ) \frac{d T ( t )}{d t}
- X_{\rm A}^{\rm neq} ( t ) \frac{d x ( t )}{d t} + \frac{d Q_{\rm A}^{\rm wst} ( t )}{d t}  , 
\end{equation}
where $S_{\rm A}^{\rm neq} ( t ) = k_{\rm B} \Lambda_{\rm A}^{\rm neq} ( t )$ is a non-equilibrium entropy,
and we have introduced the waste heat current as
\begin{eqnarray}
\label{eq:DefWstHeat}
\frac{d Q_{\rm A}^{\rm wst} ( t )}{d t} 
= - \frac{1}{\beta ( t )} \frac{ d\Sigma_{\rm A} ( t  )}{d t}.
\end{eqnarray}
From Eq.~\eqref{eq:neqLegendreP-M}, we obtain the TDL transformation between the non-equilibrium Gibbs and Helmholtz potentials as
\begin{equation}
\label{eq:NeqLegendreH-G}
F_{\rm A}^{\rm neq} ( t ) = G_{\rm A}^{\rm neq} ( t ) + x ( t ) X_{\rm A}^{\rm neq} ( t ) .
\end{equation}

Solving Eq.~\eqref{eq:neqLegendreE-M} for $U_{\rm A}^{\rm neq} ( t )$, yields the following TDL transformations:
\begin{equation}
\label{eq:NeqLegendreU-F}
U_{\rm A}^{\rm neq} ( t ) = F_{\rm A}^{\rm neq} ( t ) + T ( t ) S_{\rm A}^{\rm neq} ( t ) .
\end{equation}
From the above equations, we obtain the time derivative expressions for the enthalpy and internal energy. The results are summarized in Table~\ref{table:NeqPotential}. These results are reduced to those in Table~\ref{table:Potential} (Appendix~\ref{sec:QTP}) for the quasi-static case.  Thus, we can regard these potentials as extensions of thermodynamic potentials to a non-equilibrium regime.

Dividing both sides of Eq.~\eqref{eq:DefDLQext} by $\beta ( t )$ yields the first law of thermodynamics expressed for the internal energy as
\begin{eqnarray}
\label{eq:FirstLaw}
\frac{d U_{\rm A}^{\rm neq} ( t )}{d t} = \frac{d W_{\rm A}^{ext} ( t )}{d t} + \frac{d Q^{ext}_{\rm A} ( t )}{d t} ,
\end{eqnarray}
where
\begin{eqnarray}
\label{eq:DefQext}
\frac{d Q_{\rm A}^{ext} ( t )}{d t} = \frac{1}{\beta ( t )} \frac{d \tilde{Q}_{\rm A}^{ext} ( t )}{d t}
\end{eqnarray}
is the extensive heat current. From Eqs.~\eqref{eq:WasteHeat2} and~\eqref{eq:DefQext}, we can evaluate the  entropy production rate as
\begin{eqnarray}
\label{eq:EntropyProduction}
\frac{d \Sigma_{\rm A} ( t )}{d t} = \frac{1}{k_{\rm B}} 
\left\{ \frac{d S_{\rm A}^{\rm neq} ( t )}{d t} - \frac{1}{T ( t )} \frac{d Q_{\rm A}^{ext} ( t )}{d t} \right\} .
\end{eqnarray}
In the isothermal case, by integrating both sides of the two equilibrium states over time $t$, we obtain\cite{ST20JCP,KT22JCP1} 
\begin{eqnarray}
\Sigma_{\rm A} ( t ) = \frac{1}{k_{\rm B}} 
\left\{ \Delta S_{\rm A}^{\rm qst} - \frac{Q_{\rm A}^{ext}}{T} \right\}.
\end{eqnarray}

The non-equilibrium Gibbs energy satisfies
\begin{equation}
S^{\rm neq}_{\rm A} ( t ) = 
- \left( \frac{\partial G^{\rm neq}_{\rm A}}{\partial T ( t )} \right)_{x ( t ) , \tilde{Q}^{\rm wst}} ,
\end{equation}
\begin{equation}
X_{\rm A} ( t ) = - \left( \frac{\partial G_{\rm A}^{\rm neq}}{\partial x ( t )} \right)_{T ( t ) , \tilde{Q}^{\rm wst}} ,
\end{equation}
and
\begin{eqnarray}
\label{neqGH}
H_{\rm A} ( t ) = - T^2 ( t ) \frac{\partial}{\partial T ( t )}
\left( \frac{G^{\rm neq}_{\rm A} ( t )}{T ( t )} \right)_{x ( t ) , \tilde{Q}^{\rm wst}} .
\end{eqnarray}
Equation~\eqref{neqGH} extends the Gibbs--Helmholtz relation to a non-equilibrium regime.

\section{Numerical Demonstration}
\label{sec:Simulation}

Although the results presented in Tables~\ref{table:NeqDLPotential} and ~\ref{table:NeqPotential} hold for any non-equilibrium system consisting of subsystem and bath, the extensive variables and entropy production that appear in these relationships are system-specific, and there is no general theory on the basis of which non-equilibrium thermodynamic potentials can be obtained. However, it is possible to evaluate them as functions of intensive and extensive variables using an optimization algorithm.  Here, as a demonstration, we evaluate non-equilibrium thermodynamic potentials numerically using the thermodynamic SB model.\cite{KT24JCP1,KT24JCP3}

\subsection{Thermodynamic system-bath model}
\label{sec:model}
We employed the Ullersma--Caldeira--Leggett (or Brownian) model,\cite{Ullersma1966_1,Ullersma1966_2,CALDEIRA1983587,GRABERT1988115, Weiss2012,TW91PRA,TW92JCP} which is ideal for thermodynamic simulations because the subsystem and bath are well defined, and rigorous numerical solutions can be obtained in both classical and quantum cases under any time-dependent external perturbation. Many of the favorable features for thermodynamic investigations arise from the presence of a counter term, which allows us to include the contribution of the SB interactions in the bath.\cite{T15JCP,T06JPSJ,T20JCP,TW91PRA,TW92JCP,KT13JPCB}

By introducing multiple heat baths at different temperatures controlled by time-dependent SB coupling functions, we can investigate isothermal, isentropic, thermostatic, and entropic processes. 
The total Hamiltonian for isothermal and thermostatic processes is written as\cite{KT24JCP1,KT24JCP3}
\begin{equation}
\label{TotalHamiltonian}
\hat{H}_{\rm tot}(t) = \hat{H}_A^{0} + \hat{H}_{\rm A}'(t) +\sum_{k=0}^N  \hat{H}_{\rm IB}^k (t) ,
\end{equation}
where 
\begin{eqnarray}
\label{eq:DefSysH}
\hat{H}_{\rm A}^0= \frac{\hat{p}^2}{2 m} + U ( \hat{q} ) 
\end{eqnarray}
is the unperturbed Hamiltonian of a subsystem with mass $m$ and potential $U(\hat q)$ described by momentum $\hat{p}$ and position $\hat{q}$. The internal energy is then evaluated as $U_{\rm A}^{\rm neq} (t)=\mathrm{tr} \{ \hat{H}_{\rm A}^{0}  \hat \rho_{\rm A}(t) \}$, where $\hat{\rho}_{\rm A} ( t )$ is the reduced density operator of the subsystem.
The external perturbation is expressed as $\hat{H}_{\rm A}'(t) \equiv- x ( t ) \hat X_{\rm A}$, where $\hat X_{\rm A}$ is an operator of the subsystem coordinate [i.e., $\hat X_{\rm A}(\hat q)$], and $x ( t )$ is the thermodynamic intensive variable.  The extensive variable is evaluated as $X_{\rm A}^{\rm neq} (t)=\mathrm{tr} \{ \hat X_{\rm A} \hat \rho_{\rm A}(t) \}$.

Although the conventional SB model has been limited to the investigation of isothermal processes at constant temperature, we can extend it to describe thermostatic processes in which temperature varies with time by introducing $N$ independent heat baths, each in the thermal equilibrium state at the inverse temperature $\beta_k\equiv 1/k_{\rm B} T_k$ connected to or disconnected from subsystem A using the window function $\xi_k(t)$.\cite{KT24JCP1}  
The $k$th bath Hamiltonian is expressed as an ensemble of harmonic oscillators and is given by
\begin{equation} 
\label{B0} 
\hat{H}_{\rm IB}^k (t) \equiv   \sum_j \left\{ \frac{ ( \hat{p}^k_j )^2 }{2 m_j} + 
\frac{ m_{j}^k (\omega_{j}^k)^2}{2} \left[ \hat{x}_{j}^k -\frac{ c_{j}^k A_k \xi_k ( t ) \hat{q}} {m_{j}^k(\omega_{j}^k)^2}\right]^2  \right\},
\end{equation}
where the momentum, position, mass, and frequency of the $j$th bath oscillator are given by $\hat{p}_{j}^k$, $\hat{x}_{j}^k$, $m_{j}^k$, and $\omega_{j}^k$, respectively.
Here, we consider the situation, in which $N$ independent heat baths, each in the thermal equilibrium state 
$\exp(-\beta_k \hat{H}_{\rm I+B}^k)$ at the inverse temperature $\beta_k\equiv 1/k_{\rm B} T_k$, are connected or disconnected to subsystem A according to a control function $\xi_k(t)$. 
The bath temperature can be effectively expressed as
\begin{eqnarray}
T(t)=\sum_{k = 1}^N T_k \xi_k(t),
\label{Tt}
\end{eqnarray}
or the inverse temperature as $\beta ( t )=[k_{\rm B} T ( t )]^{-1}$.  As a spectral distribution function of the $k$th bath  
$J^k (\omega) \equiv \sum_{j } {\hbar (c_{j}^k A_k )^2}/({2m_{j}^k) \omega_{j}^k} \delta(\omega-\omega_{j}^k)$, we consider the Ohmic case described by
\begin{eqnarray}
\label{eq:ohmic-density}
J^k ( \omega ) = \frac{\hbar A_k^2 \omega}{\pi}
\end{eqnarray}
and assume that the time scale of quantum thermal fluctuations $\beta(t)\hbar/2  \pi$ is shorter than the time scale of the external perturbations $\beta(t)$ and $x (t)$.  This allows us to extend the low-temperature quantum Fokker--Planck equations (LT-QFPE)\cite{IT19JCTC} for the set of Wigner distribution functions $W_{\vec{n}}(p, q; t)$, where ${\vec{n}}$ is the index of hierarchy members, and the Kramers equation for the phase space distribution function  $W(p, q; t)$ to  the thermostatic case by introducing time-dependent Matsubara frequencies $\nu(t) = 1/\hbar \beta(t)$. 
The expressions for the thermodynamic quantum Fokker-Planck equations (T-QFPE) and thermodynamic Kramers equation (T-KE) are given in Refs.~\onlinecite{KT24JCP1,KT24JCP3}. The source codes for them 
are also provided in the supplementary material.\cite{KT24JCP3} 

While the HEOM approach to thermodynamics treats the SB interaction as part of the main system in spin-boson systems,\cite{KT15JCP,KT16JCP,ST21JPSJ,ST20JCP,KT22JCP1,KT22JCP2,KT24JCP4} this Brownian model treats it as part of a heat bath, including the counter term.\cite{KT24JCP1,KT24JCP3}
Assuming the Ohmic SDF, in the classical and high-temperature semi-classical cases, the dynamics described by this model exhibit a Markovian feature, which can be treated in the framework of stochastic thermodynamics, while in the fully quantum case, it is not only non-Markovian, but also non-factorized owing to bathentanglement. In other words, the difference between the classical and quantum results represents a deviation from Markovian thermodynamics that arises when dealing with fully quantum processes.

It should be noted that for non-Markovian processes, the SDF-based description of a thermal bath breaks down when the correlation time of the noise is longer than the time scale of the bath temperature change.\cite{KT24JCP3} In such cases, prescriptions that simply replace $\beta$ with $\beta(t)$ are not allowed, and the hierarchy members for each bath at different temperatures must be treated separately.\cite{KT24JCP4} In this way, several baths can be operated simultaneously, although this is computationally expensive.\cite{KT15JCP,KT16JCP}

\subsection{DL work and extensive variables}

We express the solution for reduced density elements under any $x(t)$ in the Wigner representation using the zeroth member of the hierarchical Wigner functions as $W (p,q,t)\equiv W_{\vec{0}}(p, q; t)$. In the classical limit $\hbar \rightarrow 0$,  ${W}(q,p;t)$ corresponds to the classical distribution function. 
We use this to define the change in DL intensive work over time, which corresponds to power and heat flow as follows:
\begin{equation}
\label{eq:DefWintSim}
\frac{d \tilde{W}_{\rm A}^{int} ( t )}{d t}
= U_{\rm A}^{\rm neq} ( t ) \frac{d \beta ( t )}{d t} - X_{\rm A}^{\rm neq} ( t ) \frac{d \tilde{x} ( t )}{d t} ,
\end{equation}
where the extensive variables in the Wigner representation at time $t$ are expressed as
\begin{eqnarray}
\label{Mneq3}
X _{\rm A}^{\rm neq}  ( t ) = \mathrm{tr}_{\rm A} \{X_{\rm A}(q)  {W} (p,q; t) \}
\end{eqnarray}
and
\begin{eqnarray}
\label{EnthalpyNumerical}
U_{\rm A}^{\rm neq}  ( t ) = \mathrm{tr}_{\rm A} \left\{ \left[ \frac{p^2}{2 m} + U ( q ) \right] 
W ( p , q ; t ) \right\} .
\end{eqnarray}
From the definition~\eqref{eq:DefWext0} of extensive work and the TDL transformations~\eqref{eq:DefWext} and~\eqref{eq:DefWQ}, the extensive heat current can be evaluated as 
\begin{multline}
\label{eq:HeatExpression}
\frac{d Q_{\rm A}^{ext} ( t )}{d t}
=\\ \mathrm{tr}_{\rm A} \left\{ \left[ \frac{p^2}{2 m} + U( q) -  x(t)X_{\rm A}(q) \right]
\frac{\partial W ( p , q ; t )}{\partial t} \right\} . 
\end{multline}

\subsection{Evaluation of the non-equilibrium Planck potential}
\label{sec:EvaluationMethod}

\begin{figure}

\includegraphics[width=8.5cm]{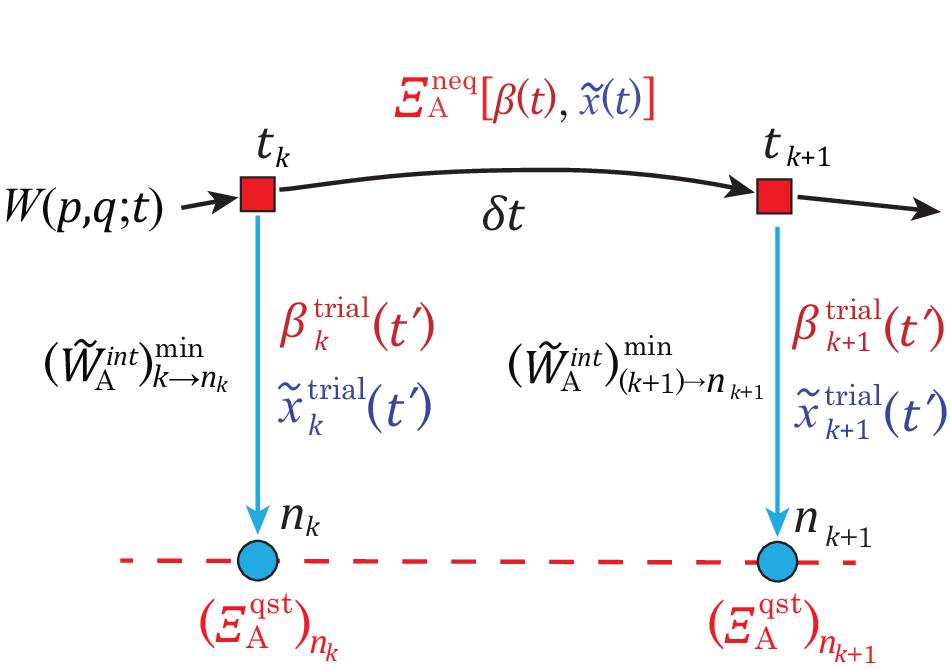}
\caption{\label{fig:Optimization} Schematic  of the evaluation of the non-equilibrium Planck potential $\Xi_{\rm A}^{\rm neq} (t)$ at $t_k$ and $t_{k+1}$. A non-equilibrium process driven by the intensive variables $\beta ( t )$ and $\tilde{x} ( t )$ is described as the non-equilibrium distribution $W(p,q;t)$ expressed by the black arrows. The equilibrium states for different sets of $\beta^{\rm qst}_{n_k}$ and $\tilde{x}_{n_k}^{\rm qst}$ are expressed by the red dashed line.  For each state $k$ to the equilibrium state $n_k$, 
we minimize $( \tilde{W}_{\rm A}^{int} )_{k \rightarrow n_k} + ( \Xi_{\rm A}^{\rm qst} )_{n_k}$ by choosing the trial functions $\beta_k^{\rm trial} (t' )$ and $\tilde{x}_k^{\rm trial} (t' )$, where $( \tilde{W}_{\rm A}^{int} )_{k \rightarrow n_k}^{\rm min}$ is the DL minimum intensive work for $k \rightarrow n_k$ and $( \Xi_{\rm A}^{\rm qst} )_{n_k}$ is the quasi-static Planck potential at the state $n_k$. We can then evaluate the non-equilibrium Planck potential at state $k$ as $( \Xi_{\rm A}^{\rm neq} )_k = ( \tilde{W}_{\rm A}^{int} )_{k \rightarrow n_k}^{\rm min} + ( \Xi_{\rm A}^{\rm neq} )_{n_k}$. }
\end{figure}

From Eq.~\eqref{eq:A2}, we obtain the following expression for the non-equilibrium Planck potential $(\Xi_{\rm A}^{\rm neq} )_k = \Xi_{\rm A}^{\rm neq} [\beta(t_k), \tilde{x}(t_k)]$ of the non-equilibrium-to-equilibrium process $k \rightarrow n_k$ as 
\begin{eqnarray}
\label{eq:PlanckEvaluation1}
( \Xi_{\rm A}^{\rm neq} )_k = ( \tilde{W}_{\rm A}^{int} )_{k \rightarrow n_k}^{\rm min} 
+ ( \Xi_{\rm A}^{\rm qst} )_{n_k} ,
\end{eqnarray} 
where $n_k$ represents the equilibrium state,  $( \tilde{W}_{\rm A}^{int} )_{k \rightarrow n_k}^{\rm min}$ is the DL minimum intensive work for $k \rightarrow n_k$, and $( \Xi_{\rm A}^{\rm qst} )_{n_k}$ is the quasi-static Planck potential at state $n_k$.  As shown in Eq.~\eqref{eq:PhiWellDefined}, the right-hand side of Eq.~\eqref{eq:PlanckEvaluation1} is independent of the choice of $n_k$; thus, to evaluate $(\Xi_{\rm A}^{\rm neq} )_k$, we do not specify $n_k$ to minimize $( \tilde{W}_{\rm A}^{int} )_{k \rightarrow n_k} + ( \Xi_{\rm A}^{\rm qst} )_{n_k}$.  

To perform numerical calculations, we express Eq.~\eqref{eq:PlanckEvaluation1} with Eq.~\eqref{eq:DiffNeqPlanck} in terms of trial functionals $\beta^{\rm trial} ( t_k , t' ) $ and $\tilde{x}^{\rm trial} ( t_k , t' )$ as
\begin{eqnarray}
\label{eq:TargetFunc}
X_{\rm A}^{\rm target} [ \beta^{\rm trial}  ( t ) , x^{\rm trial} (t) ]
= \Xi^{\rm qst}_{\rm A} ( \beta^{\rm qst}_{n_k} , x^{\rm qst}_{n_k} ) \nonumber \\
+ \int_{t}^{t + \Delta t} \left[ U_{\rm A} ( t' ) \frac{d \beta^{\rm trial}  ( t' )}{d t'} - X_{\rm A} ( t' ) \frac{d \tilde{x}^{\rm trial}  ( t' )}{d t'} \right] d t' ,
\end{eqnarray} 
where $\Xi_{\rm A}^{\rm target} [ \beta^{\rm trial} ( t ) , \tilde{x}^{\rm trial} ( t ) ]$ is the target function to be minimized and 
$\Xi_{\rm A}^{\rm qst} ( \beta_{n_k}^{\rm qst} , \tilde{x}_{n_k}^{\rm qst} )$ is the quasi-static Planck potential at $n_k$, which is also evaluated from the T-QFPE and T-KE. The DL trial functional $\tilde{x}_k^{\rm trial} ( t ) $ in Eq. \eqref{eq:TargetFunc} is evaluated from $x_k^{\rm trial} ( t )$ as $\tilde{x}_k^{\rm trial} ( t ) = \beta_k^{\rm trial} ( t ) x_k^{\rm trial} ( t )$.
 To reduce computational cost, we assume that the optimized trial functionals are constant after the characteristic time for the equilibration $\Delta t$, as $\beta_k^{\rm trial} (t_k + \Delta t )=\beta^{\rm qst}_{n_k} $ and $ x_k^{\rm trial}  ( t_k +  \Delta t )=x_{n_k}^{\rm qst}$.

As trial functionals for time $t' > t_k$, we chose the ${N_\beta}$th and ${N_x}$th Taylor expansion forms expressed as
\begin{eqnarray}
\label{eq:TrialBeta}
\beta_k^{\rm trial} (t' ) = \sum_{n = 0}^{N_\beta} \beta_k^{(n)} ( t' - t_k )^n ,
\end{eqnarray}
and
\begin{eqnarray}
\label{eq:TrialE}
x_k^{\rm trial} (t' ) = \sum_{n = 0}^{N_x} x_k^{(n )} ( t' - t_k )^n ,
\end{eqnarray}
where  $\beta_k^{(n)} $ and $x_k^{(n )}$ are the $n$th-order Taylor coefficients. Thus, the functional minimization of $\Xi_{\rm A}^{\rm target} [ \beta^{\rm trial}  ( t ) , \tilde{x}^{\rm trial} (t) ]$ becomes a multivariable functional minimization for $\beta_k^{(n)} $ and $x_k^{(n )}$. 

Then, $\Xi_{\rm A}^{\rm neq} (t)$ is evaluated as follows (see Fig.~\ref{fig:Optimization}).  First, we perform the non-equilibrium simulations for given $\beta(t)$ and $\tilde{x}(t)$ to obtain $W_{\vec{n}}(p, q; t)$ for each $t_k$.  The minimum work from the non-equilibrium state $k$ to the equilibrium state $n_k$ expressed as $( \tilde{W}_{\rm A}^{int} )_{k \rightarrow n_k}^{\rm min}$ is then evaluated by an optimization algorithm for $\beta_k^{\rm trial} (t' )$ and $\tilde{x}_k^{\rm trial} (t' )$.  From Eq.~\eqref{eq:PlanckEvaluation1}, we set this value as $( \Xi_{\rm A}^{\rm neq} )_k$. 
By repeating this procedure for different $k$ values, we obtain $\Xi_{\rm A}^{\rm neq} (t_k)$ at each step. The other potentials can be evaluated by using the TDL transformations.

\subsection{Thermostatic quantum and classical Stirling engines}
\label{sec:StirlingEngine}

The non-equilibrium thermodynamic potentials are state variables as functions of the non-equilibrium intensive and extensive variables, which are also state variables, whereas $Q_{\rm A}^{\rm wst}$ is not; in the quasi-static limit, they reduce to conventional thermodynamic potentials.
These non-equilibrium potentials are useful because they can be used to analyze thermodynamic processes using work diagrams, as in the case of equilibrium thermodynamics.
By identifyuing a thermodynamic process that minimizes entropy production, we can construct a heat machine with maximum efficiency under non-equilibrium conditions, although the efficiency is lower than the Carnot limit.\cite{KT22JCP1,KT22JCP2}

\begin{figure}

\includegraphics[width=8cm]{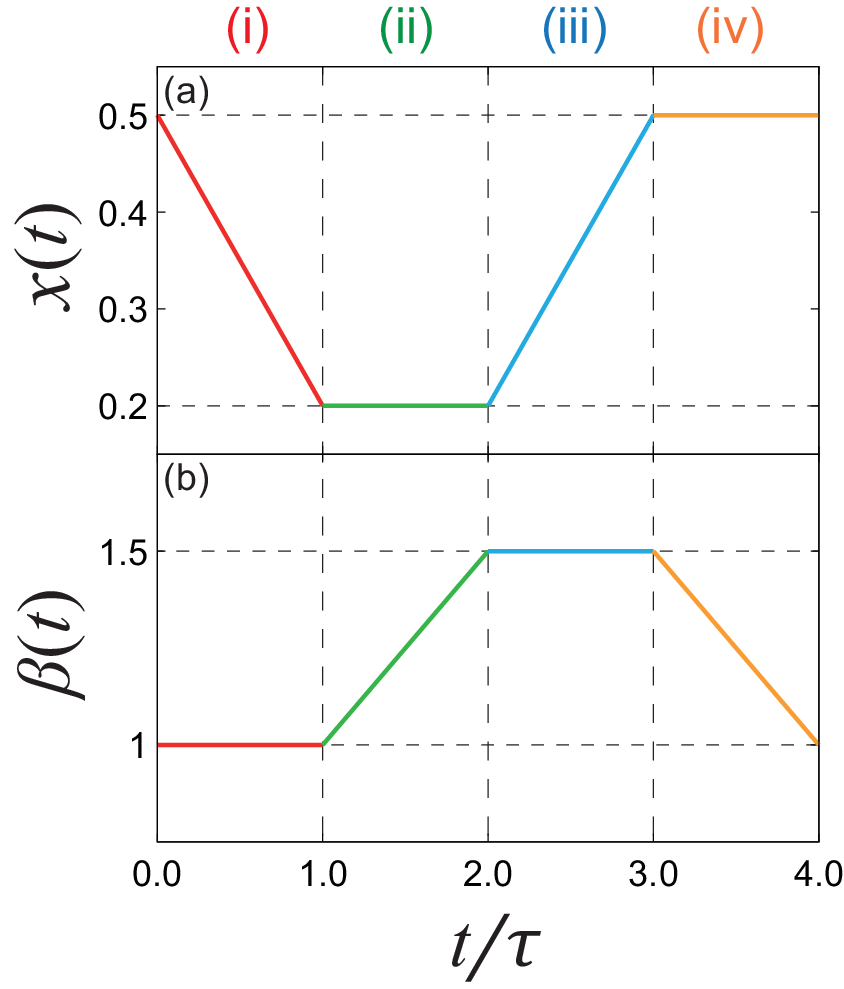}
\caption{\label{figParameters} Time profiles of (a) the electric field $x(t)$ and (b) the inverse temperature $\beta(t)$ for the thermostatic Stirling engine. The red, green, blue, and orange lines represent (i) hot isothermal, (ii) isoelectric ( hot to cold), (iii) cold isothermal, and (iv) isoelectric (cold to hot) processes, respectively.\cite{KT24JCP3} The time $t$ is normalized by the period of the cycle $\tau$.}

\end{figure}

Here, we demonstrate how to evaluate non-equilibrium potentials for the case in which $\beta(t)$ and $x(t)$ are specified. For this purpose, we consider a thermostatic classical/quantum Stirling engine\cite{KT24JCP3} consisting of four steps: (i) a hot isothermal process, (ii) a transition from a hot-to-cold thermostatic process, (iii) a cold isothermal process, and (iv) a transition from a cold-to-hot thermostatic process for an anharmonic potential system.  Since the purpose of the calculation is to illustrate how potentials are calculated, we fix the changes in $\beta(t)$ and $x(t)$ as shown in Fig.~\ref{figParameters} with $\tau = 20$ and do not optimize them to improve their efficiency.

\subsubsection{Simulation details}
\label{sec:NumericalDetails}

We performed simulations using the anharmonic potential system employed in our previous studies.\cite{KT24JCP1,KT24JCP3} Thus, we considered a quartic anharmonic potential with the external interaction described as $X_{\rm A}(\hat q)= \hat q$. The potential function in Eq.~\eqref{eq:DefSysH} is expressed as
\begin{eqnarray}
\label{SimulationSubsystemHamiltonian}
U(\hat q) = U_2 \hat{q}^2 + U_3 \hat q^3 + U_4 \hat q^4 ,
\end{eqnarray}
where the constants are given by $U_2 = 0.1$, $U_3= 0.02$, and $U_4=0.05$. 
Numerical calculations were performed to integrate the T-QFPE in the quantum cases and the T-KE in the classical cases. The detailed conditions for the numerical calculations, including the working parameters, are presented in Table \ref{table:SimDetail} and in Ref.~\onlinecite{KT24JCP1}. Source code and results for the quasi-static case are presented in Ref.~\onlinecite{KT24JCP3}.

\begin{table}[!t]
\caption{\label{table:SimDetail} Parameter values used for the simulations of the thermostatic Stirling engine.  Here, $d x$ and $d p$ are the mesh sizes for position and momentum, respectively, in  Wigner space. The integers $N$ and $K$ are the cutoff numbers used in the T-QFPE.}
\begin{ruledtabular}
\begin{tabular}{lcccccd}
& $ A $ & $ N $ & $ K $ & \multicolumn{1}{c}{$d x$}  &  \multicolumn{1}{c}{$d p$} &
\\
\hline
\multirow{2}{*}{Classical} & \multirow{2}{*}{$0.5$--$1.5$} & \multirow{2}{*}{$\cdots$} 
& \multirow{2}{*}{$\cdots$} & \multirow{2}{*}{$0.25$} & \multirow{2}{*}{$0.25$} &
\\ \\
\hline
\multirow{3}{*}{Quantum} & $0.5$ & $6$ & $2$ & $0.3$ & $0.4$ &
\\
& $1.0$ & $7$ & $2$ & $0.3$ & $0.5$ &
\\
& $1.5$ & $8$ & $2$ & $0.3$ & $0.6$ &
\end{tabular}
\end{ruledtabular}
\end{table}

To set the trial functions defined in Eqs.~\eqref{eq:TrialBeta} and~\eqref{eq:TrialE}, we chose $N_\beta = N_x = 5$. 
Then, using the Nelder--Mead method, we minimized $\Xi_{\rm A}^{\rm target} [ \beta^{\rm trial}  ( t ) , x^{\rm trial} (t) ]$ 
in Eq.~\eqref{eq:TargetFunc} with a cutoff time $\Delta t = 1.0$.  We evaluated $\Xi_{\rm A}^{\rm neq} ( t )$ in 20 steps. The value of the quasi-static Planck potential $\Xi_{\rm A}^{\rm qst} ( \beta^{\rm qst} , \tilde{x}^{\rm qst} )$ in Eq.~\eqref{eq:TargetFunc} was obtained by integrating the thermodynamic T-QFPE and the T-KE.

\subsubsection{Results}
\label{sec:Results}

\begin{figure}

\includegraphics[width=7cm]{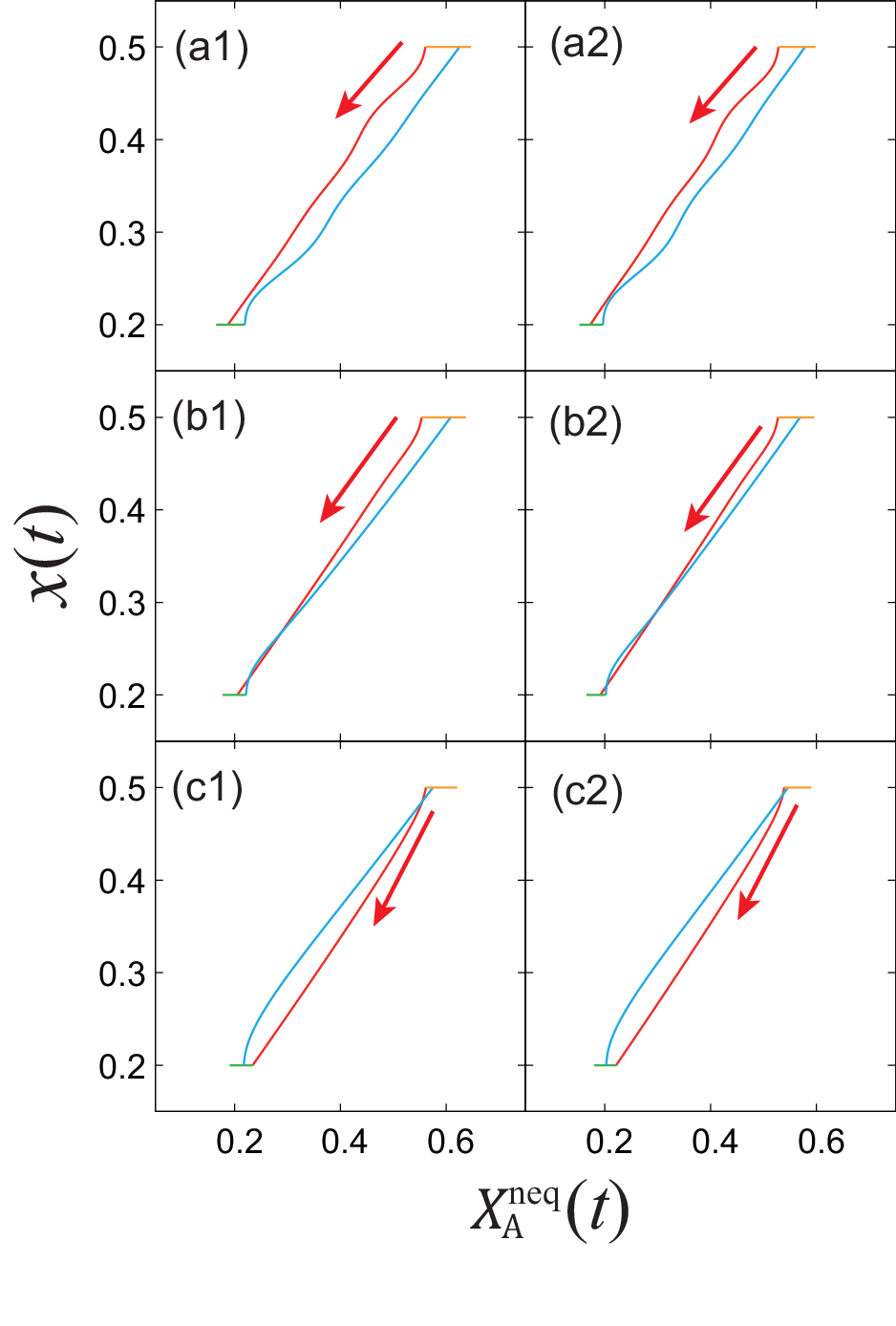}

\caption{\label{fig:EP} $x(t)$--$X_{\rm A}^{\rm neq}  ( t )$ diagrams for the thermostatic Stirling engine in the classical case (1, left column) and quantum case (2, right column) for (a) $A = 0.5$ (weak), (b) $1.0$ (intermediate), and (c) $1.5$ (strong) SB coupling strengths. In each plot, the four curves (or lines) represent (i) hot isothermal (red), (ii) hot to cold thermostatic (green), (iii) cold isothermal (blue), and (iv) cold to hot thermostatic (orange) processes, respectively.
These processes evolve in a counterclockwise manner over time in a heat engine, whereas they evolve in a clockwise manner over time in a refrigerator.}
\end{figure}

\begin{figure}

\includegraphics[width=7cm]{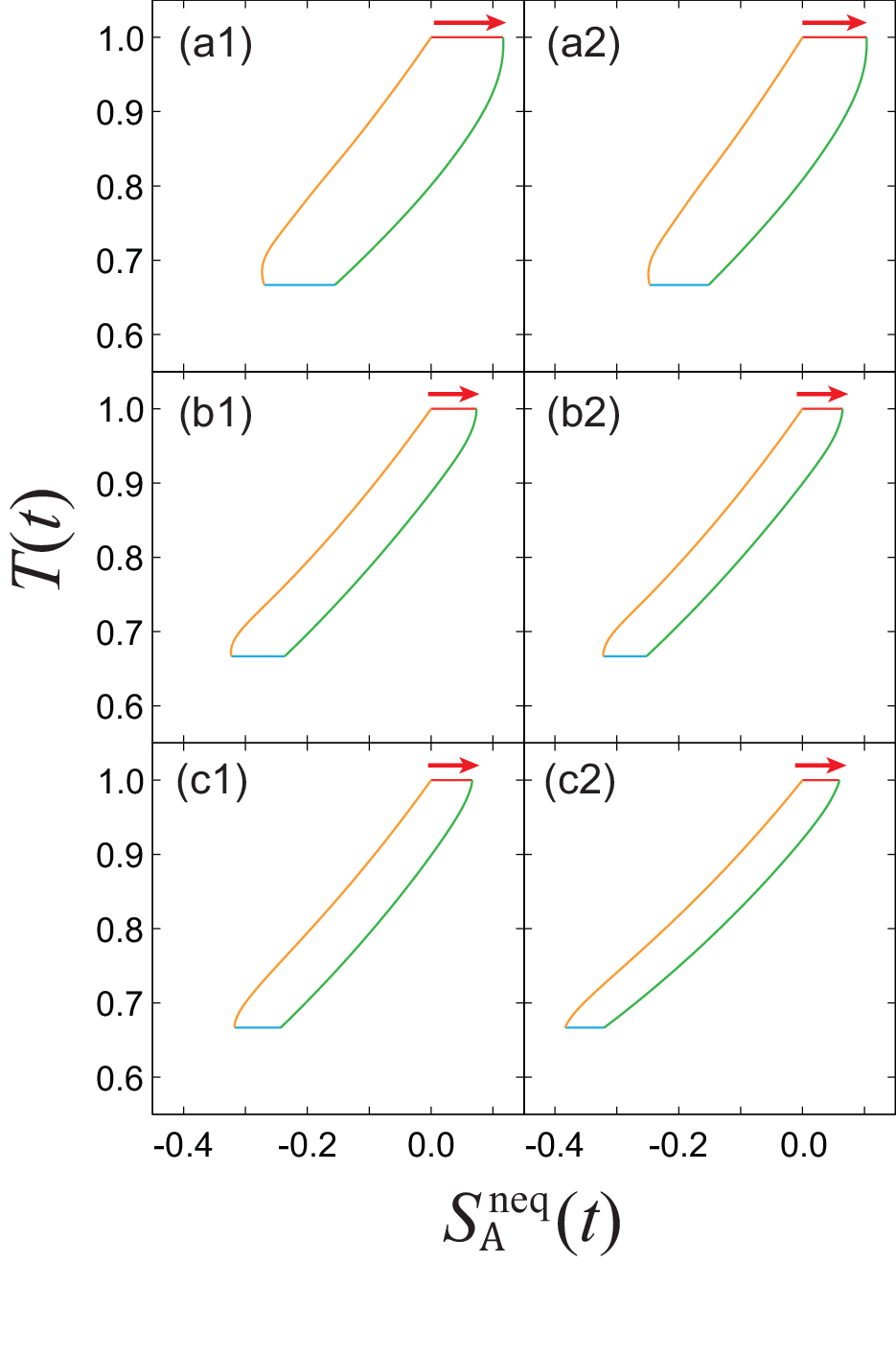}

\caption{\label{fig:TS} $T ( t )$--$S_{\rm A}^{\rm neq} ( t )$ diagrams for the thermostatic Stirling engine in the classical case (1, left column) and quantum case (2, right column) for (a) $A = 0.5$ (weak), (b) $1.0$ (intermediate), and (c) $1.5$ (strong) SB coupling strengths, respectively. Each cycle starts with a red arrow, and the four curves represent (i) hot isothermal (red), (ii) from hot to cold thermostatic (green), (iii) cold isothermal (blue), and (iv) from cold to hot thermostatic (orange) processes, respectively.}

\end{figure}

\begin{table}[!t]

\caption{\label{table:work} Work  performed in one cycle for the classical and quantum cases for different SB coupling strengths.}

\begin{ruledtabular}
\begin{tabular}{lcc}
$A$ & $W$ (classical) & $W$ (quantum)
\\
\hline
$0.5$ (weak) & $-1.44 \times 10^{-2}$  & $-1.07 \times 10^{-2}$
\\
$1.0$ (intermediate) & $-5.77 \times 10^{-3}$ & $-3.09 \times 10^{-3}$
\\
$1.5$ (strong) & $1.01 \times 10^{-2}$ & $1.10 \times 10^{-2}$
\end{tabular}
\end{ruledtabular}

\end{table}

Extensive variables $X_{\rm A}^{\rm neq} ( t )$ and $U_{\rm A}^{\rm neq} ( t )$ were obtained from Eqs.~\eqref{Mneq3} and~\eqref{EnthalpyNumerical}, respectively.  The non-equilibrium entropy expressed as 
$S_{\rm A}^{\rm neq} ( t ) = k_{\rm B} \Lambda_{\rm A}^{\rm neq} ( t )$ was then obtained from $X_{\rm A}^{\rm neq} ( t )$, and $U_{\rm A}^{\rm neq} ( t )$ and $\Xi_{\rm A}^{\rm neq} ( t )$ were evaluated according to the procedure described in Sec.~\ref{sec:EvaluationMethod} using the TDL transformations given by Eqs. \eqref{eq:neqLegendreP-M} and \eqref{eq:neqLegendreE-M}.  
The results are depicted as  non-equilibrium $x(t)$--$X_{\rm A}^{\rm neq} ( t )$ and  $T ( t )$--$S_{\rm A}^{\rm neq} ( t )$ diagrams, whose cycle trajectories are closed because the cycle is stationary and because $T ( t )$, $S_{\rm A}^{\rm neq} ( t )$, $x ( t )$, and $X_{\rm A}^{\rm neq} ( t )$ are state variables.

We first present the $x(t)$--$X_{\rm A}^{\rm neq} ( t )$ diagrams for  weak ($A = 0.5$), intermediate ($A = 1.0$), and strong ($A = 1.5$) SB coupling strengths in Fig.~\ref{fig:EP}. 
The results for the quasi-static case are presented in Fig. 5 in Ref.~\onlinecite{KT24JCP3}.

The area enclosed by each diagram corresponds to positive work when evolving in the counterclockwise direction, whereas negative work corresponds to clockwise evolution. 
The work performed in one cycle is presented in Table~\ref{table:work}. In each diagram in Fig.~\ref{fig:EP}, the red and blue horizontal lines appear because a time delay exists in the change of $X_{\rm A}^{\rm neq} ( t )$ with respect to $x(t)$ for the heat bath to take effect. 

As shown in Fig.~\ref{fig:EP} and Table~\ref{table:work}, the larger the SB coupling, the smaller the quantum effect because it is suppressed by relaxation.\cite{IT19JCTC} 
Thus, when the coupling is large, dissipation dominates, and the system becomes a refrigerator (or damper). This is because the time scales of fluctuation and dissipation are differ: in the non-equilibrium case, dissipation dominates when SB coupling is large, whereas in the quasi-static case (Fig. 5 in Ref.~\onlinecite{KT24JCP3}), fluctuations and dissipation are balanced regardless of the coupling strength. This is a distinct difference from the Carnot case; the efficiency reaches a maximum in the intermediate SB coupling region, where $A$ is neither large nor small.\cite{KT22JCP2}  

Next, we present the $T ( t )$--$S_{\rm A}^{\rm neq} ( t )$ diagrams in Fig.~\ref{fig:TS}. 
The results for the quasi-static case are presented in Fig. 6 in Ref.~\onlinecite{KT24JCP3}.
The area enclosed by the clockwise curve represents the difference between the extensive heat and waste heat, $Q_{\rm A}^{ext}  - Q_{\rm A}^{\rm wst}$, per cycle.  In the quasi-static case (Fig. 6 in Ref.~\onlinecite{KT24JCP3}), the area agree with the extensive heat per cycle because $Q^{\rm wst}$ becomes zero.  In the thermostatic processes (green and orange curves), when SB coupling is small, a time delay is observed in the change in $S_{\rm A}^{\rm neq} ( t )$ with respect to $T(t)$.  This occurs because it takes time for the system to be excited when thermal fluctuations are small, whereas the delay is almost negligible when SB coupling is strong.

\begin{figure}

\includegraphics[width=7cm]{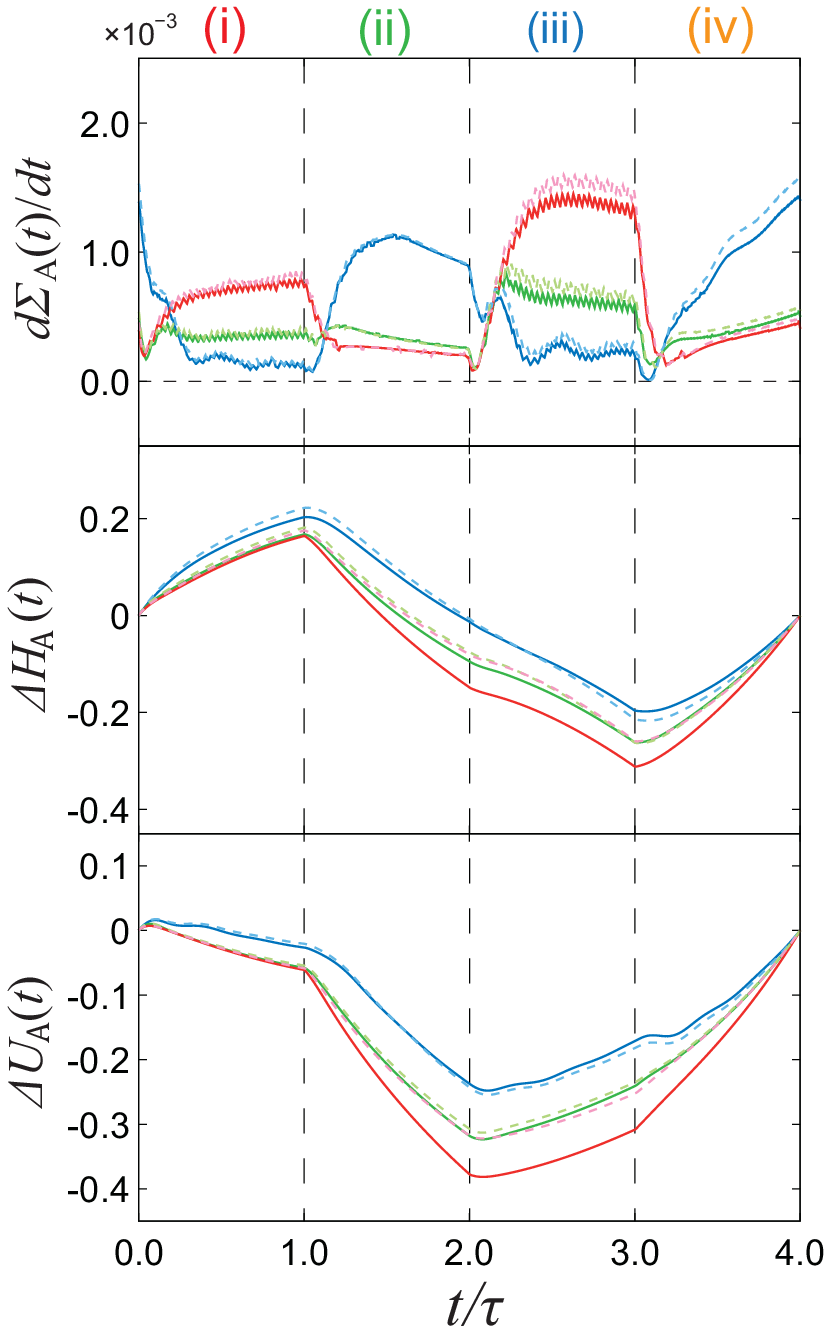}

\caption{\label{fig:Result2} (a) Entropy production rate $d \Sigma_{\rm A} ( t ) / d t$, (b) the change of enthalpy $\Delta H_{\rm A}(t)$, and (c) internal energy $\Delta U_{\rm A}(t)$, as  functions of $t$. The blue, green, and red curves represent the weak ($A = 0.5$), intermediate ($A = 1.0$), and strong ($A = 1.5$) SB coupling cases, respectively. The solid and dashed curves represent the quantum and classical results, respectively. 
The plots of the entropy production rate were time-averaged over 21 steps each using the raw data.}

\end{figure}

In Fig.~\ref{fig:Result2}, we show the (time-averaged) entropy production rate, enthalpy, and internal energy as  functions of $t$. As we show in Sec.~\ref{sec:Thermodynamics}, the entropy production rate is always positive, and becomes large in the isothermal case for larger SB coupling because of strong dissipation, whereas it becomes small in the thermostatic processes for the case of weak SB coupling  because  fluctuations are suppressed. 

Using the graph of the entropy production rate, we can improve the cycle efficiency. For instance, in the case of strong coupling, because a large entropy production rate occurs in isothermal processes, increasing the duration of the isothermal processes reduces the entropy production in the cycle.

For weak SB coupling, both the enthalpy and internal energy in the classical case agree with those in the quantum case, whereas for strong SB coupling, they disagree because of bathentanglement.\cite{T20JCP}

\section{Conclusions}
\label{sec:conclude}

The virtue of thermodynamics lies in its ability to describe, in a system-independent manner, macroscopic thermal phenomena resulting from complex microscopic interactions as changes in thermodynamic potentials that are described as interrelated intensive and extensive variables via Legendre transformations. 

We developed the laws of thermodynamics as applied to work in a system-independent manner, based on the principle of DL minimum work. Subsequently, we have developed a non-equilibrium thermodynamic theory that describes the non-equilibrium Massieu--Planck potentials and Helmholtz--Gibbs potentials in time derivative form in terms of non-equilibrium extensive and intensive variables, which are state variables along with entropy production, which is not a state variable. 
Our results are summarized in Tables~\ref{table:NeqDLPotential} and~\ref{table:NeqPotential} and are consistent with traditional thermodynamics in a quasi-static case, as presented in Tables~\ref{table:DLPotential} and~\ref {table:Potential}.

These results have been validated for thermostatic quantum and classical Stirling engines through numerical simulations based on the thermodynamic SB model, which can describe both isothermal and thermostatic processes.
An optimization algorithm was used to evaluate non-equilibrium thermodynamic potentials.  Work diagrams in non-equilibrium regimes were presented and analyzed.  

In this paper, to perform numerical calculations, the external field was chosen as the intensive variable and the polarization as the extensive variable.  In a Brownian model defined by a potential function, it is also possible to select pressure and volume as intensive and extensive variables, as the partition function was evaluated for an ideal gas.

Although our theory in the non-equilibrium regime is model-specific, it is possible to test it even in real systems by using optimization algorithms. By designing processes to reduce entropy production, efficient heat engines can be developed for non-equilibrium processes.

In engineering, the effective energy from non-equilibrium to equilibrium is referred to as exergy.\cite{Rant1956Exergie} The non-equilibrium thermodynamic potentials introduced can be regarded as a generalization of this concept. This theory provides a methodology to systematically evaluate and improve it, which should be useful from the perspective of Sustainable Development Goals (SDGs).

\section*{Acknowledgments}
The authors thank Yoshi Oono for his critical comments on the definition and terminology of thermodynamic variables, particularly extensive variables.
Y.T. was supported by JSPS KAKENHI (Grant No.~B21H01884).
S.K. was supported by a JST fellowship, the establishment of university fellowships towards the creation of science technology innovation (Grant No.~JPMJFS2123), and by Grant-in-Aid for JSPS Fellows (Grant No.~24KJ1373).

\section*{Author declarations}
\subsection*{Conflict of Interest}
The authors have no conflicts to disclose.

\section*{Author Contributions}
{\bf Shoki Koyanagi}: Formal analysis (lead); Investigation (equal); Methodology (equal);
Software (lead); Writing - original draft (equal). {\bf Yoshitaka Tanimura}: Conceptualization (lead); Formal analysis (supporting); Funding acquisitiojn (lead); 
 Investigation (equal); Methodology (equal); Writing - review and editing (lead).

\section*{Data availability}
The data that support the findings of this study are available from the corresponding authors upon reasonable request.

\appendix

\section{Quasi-static thermodynamic potentials}
\label{sec:QTP}

\begin{table*}[!t]
\caption{\label{table:DLPotential}  Total differential expressions for the quasi-static (qst.) entropic potentials as functions of the intensive variables $\beta^{\rm qst}(t)$ and  $\tilde{x}^{\rm qst}(t)$ and the extensive variables $U_{\rm A}^{\rm qst}(t)$ and  ${X}_{\rm A}^{\rm qst}(t)$.\ Entropy has two definitions, depending on whether the work variable is intensive or extensive. Of these DL entropies, the commonly used one, which we call Massieu entropy (M-entropy),  involves only extensive variables and is denoted by $\Lambda_{\rm A}^{\rm qst}[U_{\rm A}^{\rm qst}, {X}_{\rm A}^{\rm qst} ]$, whereas the less widely used one, which we call Planck entropy (P-entropy),  is denoted by $\Gamma_{\rm A}^{\rm qst} [ U_{\rm A}^{\rm qst}, \tilde{x}^{\rm qst} ]$. Whereas the enthalpy $H_{\rm A}^{\rm qst}(t)$ was chosen as the natural variable in Ref.~\onlinecite{KT24JCP1}, here we  chose the internal energy $U_{\rm A}^{\rm qst}$ instead.  Each potential is related to the others by the Legendre transformations  shown in the final column.}
\begin{ruledtabular}
\begin{tabular}{llcc}
Qst. Potential & Differential form & Natural var. & Legendre transformation 
\\
\hline
Massieu & 
$d \Phi_{\rm A}^{\rm qst}  = - U_{\rm A}^{\rm qst} d \beta^{\rm qst}  - \tilde{x}^{\rm qst} d {X}_{\rm A}^{\rm qst}$ 
& $\beta^{\rm qst} , {X}_{\rm A}^{\rm qst}$ & $\cdots$ 
\\
Planck  & 
$d \Xi_{\rm A}^{\rm qst}  = - U_{\rm A}^{\rm qst} d \beta^{\rm qst} + {X}_{\rm A}^{\rm qst} d \tilde{x}^{\rm qst}$
& $\beta^{\rm qst} , \tilde{x}^{\rm qst}$ & $\Xi_{\rm A}^{\rm qst} = \Phi_{\rm A}^{\rm qst} + \tilde{x}^{\rm qst}{X}_{\rm A}^{\rm qst}$
\\
M-Entropy 
& $d \Lambda_{\rm A} ^{\rm qst} = \beta^{\rm qst} d U_{\rm A}^{\rm qst} - \tilde{x}^{\rm qst} d X_{\rm A}^{\rm qst}$
& $U_{\rm A}^{\rm qst} , X_{\rm A}^{\rm qst}$
& $\Lambda_{\rm A} ^{\rm qst} = \Xi_{\rm A} ^{\rm qst} + \beta^{\rm qst} U^{\rm qst}_{\rm A} $
\\
P-Entropy 
& $d \Gamma_{\rm A}^{\rm qst}  = \beta^{\rm qst} d U_{\rm A}^{\rm qst} + X_{\rm A}^{\rm qst} d \tilde{x}^{\rm qst}$
& $U_{\rm A}^{\rm qst} , \tilde{x}^{\rm qst}$
& $\Gamma_{\rm A}^{\rm qst}  =  \Phi_{\rm A}^{\rm qst} + \beta^{\rm qst} U^{\rm qst}_{\rm A} $
\end{tabular}
\end{ruledtabular}
\end{table*}

We introduce the DL intensive heat defined as
\begin{eqnarray}
\label{eq:DefWQint}
\frac{d \tilde{Q}_{\rm A}^{int}  ( t )}{d t} \equiv \beta ( t ) \frac{d U_{\rm A} ( t )}{d t} +
 X_{\rm A} ( t ) \frac{d \tilde{x}( t ) }{d t},
\end{eqnarray}
which satisfies the Legendre transformation 
\begin{eqnarray}
\label{eq:DefWQintLe}
\frac{d \tilde{Q}_{\rm A}^{int} ( t )}{d t} = \frac{d \tilde{Q}^{ext}_{\rm A} ( t )}{d t} + \frac{d}{d t} \left[ \tilde{x} ( t ) X_{\rm A} ( t ) \right].
\end{eqnarray}
Extensive and intensive work and heat $\tilde{W}_{\rm A}^{ext} ( t )$, $\tilde{W}_{\rm A}^{int} ( t )$, $\tilde{Q}_{\rm A}^{ext} ( t )$, and $\tilde{Q}_{\rm A}^{int} ( t )$
are interrelated via Legendre transformations~\eqref{eq:DefWext},~\eqref{eq:DefWQ}, and~\eqref{eq:DefWQintLe}. Therefore,  we obtain the inequalities~\eqref{eq:MinimumExtWork0},~\eqref{eq:MinimumintQ0},
\begin{eqnarray}
\label{eq:MinimumPrinciple0}
\tilde{W}_{\rm A}^{int} \geq - \Delta \Xi_{\rm A}^{\rm qst}, 
\end{eqnarray}
and 
\begin{eqnarray}
\label{eq:MaximumDLHeat0}
\tilde{Q}_{\rm A}^{int}  \leq\Delta \Gamma_{\rm A}^{\rm qst}, 
\end{eqnarray}
where $\Gamma_{\rm A}^{\rm qst}$ is the Planck entropy (P-entropy). These are all expressions of the second law of thermodynamics. 
The total differential forms of the Massieu--Planck potentials are presented in Table~\ref{table:DLPotential}.

The Helmholtz--Gibbs potentials can be obtained from the Massieu--Planck potentials using the definitions $F^{\rm qst}_{\rm A} ( t ) = - \Phi_{\rm A}^{\rm qst} ( t ) / \beta^{\rm qst}(t)$ and $G^{\rm qst}_{\rm A} ( t ) = - \Xi_{\rm A}^{\rm qst} ( t ) / \beta ^{\rm qst}(t)$.  
From Eq.~\eqref{eq:DefWext}, we obtain
\begin{equation}
\label{eq:qstLegendreH-G}
F_{\rm A}^{\rm qst} ( t ) = G_{\rm A}^{\rm qst} ( t ) + x ^{\rm qst}( t ) X_{\rm A}^{\rm qst} ( t ) .
\end{equation}
Accordingly, from Eq.~\eqref{eq:DefWQ}, we have
\begin{equation}
\label{eq:qstLegendreU-F}
U_{\rm A}^{\rm qst} ( t ) = F_{\rm A}^{\rm qst} ( t ) + T ^{\rm qst}( t ) S_{\rm A}^{\rm qst} ( t ) ,
\end{equation}
where we have used $d\beta^{\rm qst}(t)/dt= - ( 1 / k_{\rm B} [T^{\rm qst}(t) ]^2 ) d T^{\rm qst} ( t ) / d t$ and 
$S_{\rm A}^{\rm qst} ( t ) = k_{\rm B} \Lambda_{\rm A}^{\rm qst} ( t )$.

The total differential forms of the  Helmholtz--Gibbs potentials are presented in Table~\ref{table:Potential}.

\begin{table*}[!t]
\caption{\label{table:Potential} Total differential expressions for the quasi-static (qst.)  thermodynamic potentials as  functions of intensive variables $T^{\rm qst}(t)$ and  $x^{\rm qst}(t)$ and extensive variables $S_{\rm A}^{\rm qst}(t)$ and  ${X}_{\rm A}^{\rm qst}(t)$. The potentials are related through the Legendre transformations shown in the final column}.\cite{KT24JCP1}
\begin{ruledtabular}
\begin{tabular}{llcc}
Qst.  potential & Differential form & Natural var. & Legendre transformation 
\\
\hline
Helmholtz 
& $d F_{\rm A} ^{\rm qst} = - S_{\rm A} ^{\rm qst} d T^{\rm qst} + x^{\rm qst} d X_{\rm A}^{\rm qst}$
& $T^{\rm qst} , X_{\rm A}^{\rm qst}$
& $\cdots$ 
\\
Gibbs  
& $d G_{\rm A} ^{\rm qst} = - S_{\rm A} ^{\rm qst} d T^{\rm qst} - X_{\rm A}^{\rm qst} d x^{\rm qst}$
& $T^{\rm qst} , x^{\rm qst}$ 
& $G_{\rm A}^{\rm qst} = F_{\rm A}^{\rm qst} - x^{\rm qst} X_{\rm A}^{\rm qst}$ 
\\
Internal 
& $d U_{\rm A}^{\rm qst} = T^{\rm qst} d S_{\rm A}^{\rm qst} + x^{\rm qst} d X_{\rm A}^{\rm qst}$
& $S_{\rm A}^{\rm qst} , X_{\rm A}^{\rm qst}$
& $U_{\rm A}^{\rm qst} =  F_{\rm A}^{\rm qst} + T^{\rm qst} S_{\rm A}^{\rm qst}$
\\
Enthalpy
& $d H_{\rm A}^{\rm qst} = T^{\rm qst} d S_{\rm A}^{\rm qst} - X_{\rm A}^{\rm qst} d x^{\rm qst}$
& $S_{\rm A}^{\rm qst} , x^{\rm qst}$
& $H^{\rm qst}_{\rm A} = G_{\rm A} ^{\rm qst} + T^{\rm qst} S_{\rm A}^{\rm qst}$
\end{tabular}
\end{ruledtabular}
\end{table*}

\section{DL non-equilibrium minimum work principle}
\label{sec:DNEMWP}

\begin{figure}
\includegraphics[width=8cm]{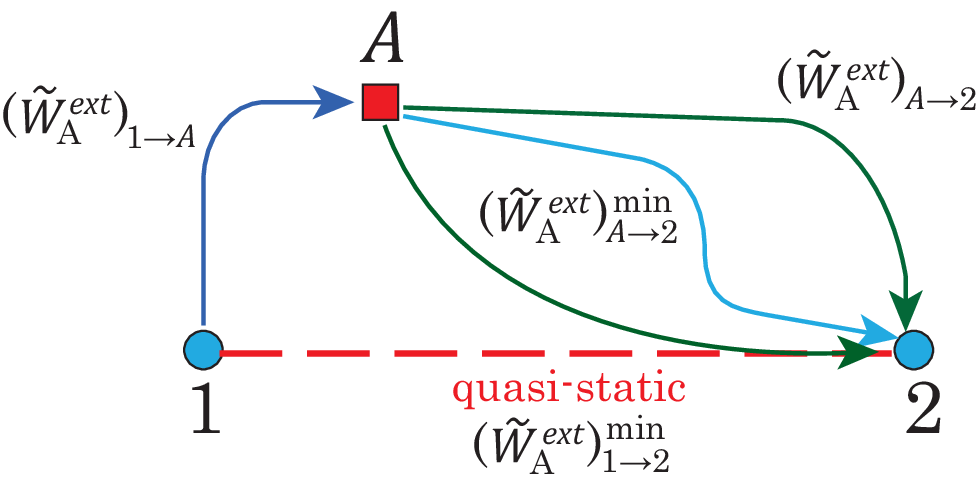}
\caption{\label{fig:Rel1} Schematic  to derive Eq.~\eqref{eq:A2}.  The blue circles 1 and 2 represent the equilibrium states, and the red square A represents the non-equilibrium state.  The quasi-static work performed for $1 \rightarrow 2$ (red dashed line) is denoted by $(\tilde{W}_{\rm A}^{ext})_{1 \rightarrow 2}^{\rm min}$, while the non-equilibrium work from 1 to A (dark blue curve) is denoted by $(\tilde{W}_{\rm A}^{ext})_{1 \rightarrow A}$.  When $(\tilde{W}_{\rm A}^{ext})_{1 \rightarrow A}$ is fixed, the value of $( \tilde{W}_{\rm A}^{ext})_{A \rightarrow 2}$  depends on the path (green and light blue curves).  From the second law of thermodynamics, we have $( W_{\rm A}^{ext} )_{1 \rightarrow A}+ ( \tilde{W}_{\rm A}^{ext} )_{\rm A \rightarrow 2} \geq - 
( \Delta \Phi_{\rm A}^{\rm qst} )_{1 \rightarrow 2}$.  This indicates a lower bound of work for ${A \rightarrow 2}$
denoted by $(\tilde{W}_{\rm A}^{ext} )_{A \rightarrow 2}^{\rm min} $ (light blue curve) and given by Eq.~\eqref{eq:A2}. 
 }
\end{figure}

Here, we derive fundamental equations to develop the DL non-equilibrium minimum work principle.  
Consider two equilibrium states 1 and 2, which are connected by a quasi-static process (see Fig.~\ref{fig:Rel1}). From the principle of DL minimum work [Eq.~\eqref{eq:MinimumExtWork0}], the work performed in this process is equivalent to the difference in the Massieu potentials: $(\tilde{W}_{\rm A}^{ext})_{1 \rightarrow 2}^{\rm min} = - (\Delta \Phi_{\rm A}^{\rm qst} )_{1 \rightarrow 2} \equiv (\Phi_{\rm A}^{\rm qst})_1 -(\Phi_{\rm A}^{\rm qst})_2 $, where ${n \rightarrow n'}$ represents the transition from any state $n$ to $n'$. Separately, we consider the non-equilibrium state $A$ and introduce the non-equilibrium process $1 \rightarrow A$ (dark blue curve in Fig.~\ref{fig:Rel1}), and any process from $A$ to 2 (green and light blue curves).  For each of these processes,  the work performed is denoted by $(\tilde{W}_{\rm A}^{ext})_{1 \rightarrow A}$ and $(\tilde{W}_{\rm A}^{ext})_{A \rightarrow 2}$, respectively.  For fixed  $(\tilde{W}_{\rm A}^{ext})_{1 \rightarrow A}$, the value of $(\tilde{W}_{\rm A}^{ext})_{A \rightarrow 2}$ changes depending on path.

From the second law of thermodynamics [Eq.~\eqref{eq:MinimumExtWork0}], we obtain
$( \tilde{W}_{\rm A}^{ext} )_{1 \rightarrow A}+ ( \tilde{W}_{\rm A}^{ext} )_{A \rightarrow 2} > - ( \Delta \Phi_{\rm A}^{\rm qst} )_{1 \rightarrow 2}.$  This indicates that $( \tilde{W}_{\rm A}^{ext} )_{A \rightarrow 2}$ has a lower bound for fixed $( \tilde{W}_{\rm A}^{ext} )_{1 \rightarrow A}$; otherwise, the inequality is violated, for example, as follows: 
$- \infty > - (\Delta \Phi_{\rm A}^{\rm qst} )_{1 \rightarrow 2} -( \tilde{W}_{\rm A}^{ext} )_{1 \rightarrow A}$. 
Using the lower bound of this work, we introduce the non-equilibrium Massieu potential  $\Phi_{\rm A}^{\rm neq}$ defined as
\begin{eqnarray}
\label{eq:A2}
( \tilde{W}_{\rm A}^{ext} )_{A \rightarrow 2}^{\rm min} =  -[ (\Phi_{\rm A}^{\rm qst})_2 - (\Phi_{\rm A}^{\rm neq})_A ].
\end{eqnarray}

\begin{figure}
\includegraphics[width=8cm]{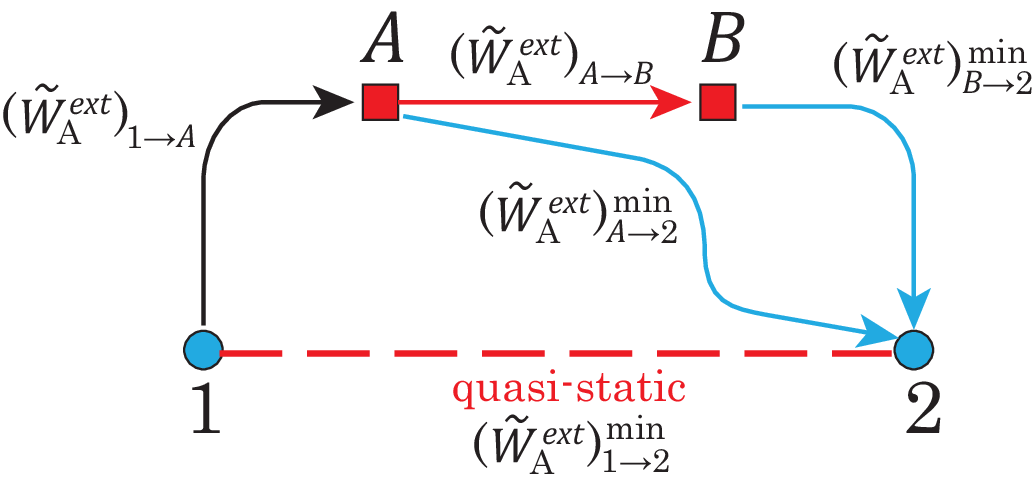}
\caption{\label{fig:Rel2}Schematic for deriving the principle of DL non-equilibrium minimum work expressed by Eq.~\eqref{eq:AB}. We consider two non-equilibrium states $A$ and $B$ (red square)  that are connected by a non-equilibrium process described by the red arrow for $A \rightarrow B$. The light blue curves $A \rightarrow 2$ and  
$B \rightarrow 2$ represent  non-equilibrium-to-equilibrium minimum work paths, whose work is denoted by 
$(\tilde{W}_{\rm A}^{ext} )_{A \rightarrow 2}^{\rm min}$ and $(\tilde{W}_{\rm A}^{ext} )_{B \rightarrow 2}^{\rm min}$, respectively. }
\end{figure}

By introducing a second non-equilibrium state $B$ on the pathway $A \rightarrow 2$ as depicted in Fig.~\ref{fig:Rel2}, we now discuss the non-equilibrium transition ${A \rightarrow B}$ (red arrow). We consider the non-equilibrium-to-equilibrium minimum work from $B$ to 2 (light blue curve), expressed as
\begin{eqnarray}
\label{eq:B2}
( \tilde{W}_{\rm A}^{ext} )_{B \rightarrow 2}^{\rm min} =  -[ (\Phi_{\rm A}^{\rm qst})_2 - (\Phi_{\rm A}^{\rm neq})_B ].
\end{eqnarray}
From the inequality $( \tilde{W}_{\rm A}^{ext} )_{A \rightarrow 2} \geq - [ ( \Phi_{\rm A}^{\rm qst} )_2 - ( \Phi_{\rm A}^{\rm neq} )_A ]$ and Eq.~\eqref{eq:B2}, we obtain
\begin{eqnarray}
\label{eq:AB}
( \tilde{W}_{\rm A}^{ext} )_{A \rightarrow B} \geq  - (\Delta \Phi_{\rm A}^{\rm neq})_{A \rightarrow B},
\end{eqnarray}
where $( \tilde{W}_{\rm A}^{ext} )_{A \rightarrow B} \equiv ( \tilde{W}_{\rm A}^{ext} )_{A \rightarrow 2} - ( \tilde{W}_{\rm A}^{ext} )_{B \rightarrow 2}^{\rm min}$. Thus,  equality holds in~\eqref{eq:AB}  when $B$ in on the minimal pathway $( A \rightarrow 2 )^{\rm min}$.

Because a non-equilibrium process changes as a function of time, it is more convenient to use the time derivative form of the above. Thus, for the process from $A$ to $B$ at times $t$ and $t + d t$, where $d t$ is infinitesimal time, we obtain the following inequality:
\begin{eqnarray}
\label{eq:NeqDLExtWorkIneq0}
\frac{d \tilde{W}_{\rm A}^{ext} ( t )}{d t} \geq - \frac{d \Phi_{\rm A}^{\rm neq} ( t )}{d t} .
\end{eqnarray}
This inequality extends the principle of minimum work to a non-equilibrium regime. 

\begin{figure}

\includegraphics[width=7cm]{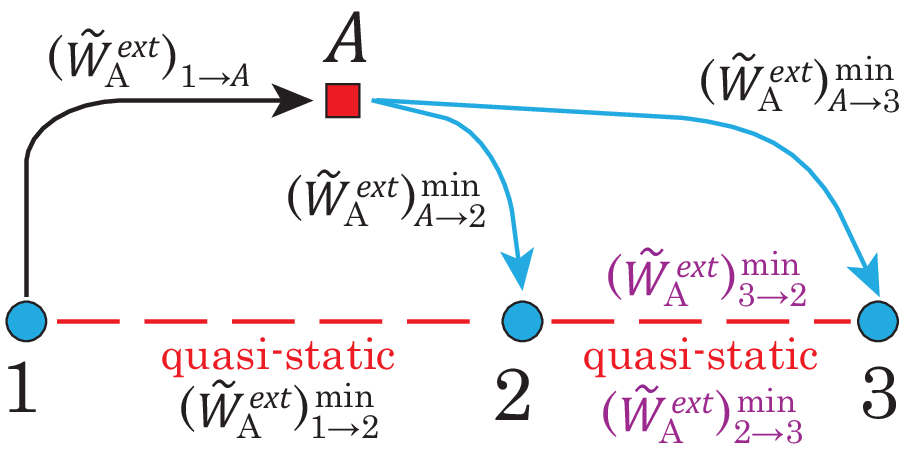}
\caption{\label{fig:Rel3} Schematic  showing that the non-equilibrium Massieu potential can be evaluated from any equilibrium state 3 instead of 2. To illustrate this, we consider the non-equilibrium-to-equilibrium path $A \rightarrow 3$, where we introduce a third equilibrium state denoted by 3 (blue circle). The work between the two equilibrium states  is denoted by  $( W_{\rm A}^{ext} )_{2 \rightarrow 3}^{\rm qst}$ and $( W_{\rm A}^{ext} )_{3 \rightarrow 2}^{\rm qst}$, respectively.}
\end{figure}

Although we have derived Eqs.~\eqref{eq:AB} and~\eqref{eq:NeqDLExtWorkIneq0} for the specific equilibrium state 2, the value 
$(\Phi_{\rm A}^{\rm neq})_{A}$ is the same for any quasi-equilibrium state along $\beta^{\rm qst}(t)$ and $x^{\rm qst}(t)$.
To illustrate this, we introduce a third equilibrium state 3 and consider the transition $A \rightarrow 3$ (see Fig.~\ref{fig:Rel3}).  Because $( \tilde{W}_{\rm A}^{ext} )_{A \rightarrow 2}^{\rm min} + ( \tilde{W}_{\rm A}^{ext} )_{2 \rightarrow 3}^{\rm qst} \geq ( \tilde{W}_{\rm A}^{ext} )_{A \rightarrow 3}^{\rm min}$ for $A \rightarrow 2 \rightarrow 3$ and
$( \tilde{W}_{\rm A}^{ext} )_{A \rightarrow 3}^{\rm min} + ( \tilde{W}_{\rm A}^{ext} )_{3 \rightarrow 2}^{\rm qst} \geq ( \tilde{W}_{\rm A}^{ext} )_{A \rightarrow 2}^{\rm min}$ for $A \rightarrow 3 \rightarrow 2$,  we have 
\begin{eqnarray}
\label{eq:PhiWellDefined}
( W_{\rm A}^{ext} )_{A \rightarrow 2}^{\rm min} + ( \Phi_{\rm A}^{\rm neq} )_2
= ( W_{\rm A}^{ext} )_{A \rightarrow 3}^{\rm min} + ( \Phi_{\rm A}^{\rm neq} )_3 ,
\end{eqnarray}
where we have used $( W_{\rm A}^{ext} )_{2 \rightarrow 3}^{\rm qst} = ( \Phi_{\rm A}^{\rm qst} )_3 - ( \Phi_{\rm A}^{\rm qst} )_2$. From Eq.~\eqref{eq:A2}, this indicates that the non-equilibrium Massieu potential is independent of the choice of the equilibrium state 2.

Because the other potentials are interrelated through TDL transformations, we obtain
\begin{eqnarray}
\label{eq:NeqDLintWorkIneq0}
\frac{d \tilde{W}_{\rm A}^{int} ( t )}{d t} \geq - \frac{d \Xi_{\rm A}^{\rm neq} ( t )}{d t},
\end{eqnarray}
\begin{eqnarray}
\label{eq:NeqDLExtQIneq0}
\frac{d \tilde{Q}_{\rm A}^{ext} ( t )}{d t} \leq \frac{d \Lambda_{\rm A}^{\rm neq} ( t )}{d t},
\end{eqnarray}
and
\begin{eqnarray}
\label{eq:NeqDLIntQIneq0}
\frac{d \tilde{Q}_{\rm A}^{int} ( t )}{d t} \leq \frac{d \Gamma_{\rm A}^{\rm neq} ( t )}{d t}.
\end{eqnarray}

For an isothermal case, from the definitions $F^{\rm neq}_{\rm A} ( t ) = - \Phi_{\rm A}^{\rm neq} ( t ) / \beta$ and $G^{\rm neq}_{\rm A} ( t ) = - \Xi_{\rm A}^{\rm neq} ( t ) / \beta$, we have
\begin{eqnarray}
\label{eq:NeqExtWorkF}
\frac{d W_{\rm A}^{ext} ( t )}{d t} \geq \frac{d F_{\rm A}^{\rm neq} ( t )}{d t} 
\end{eqnarray}
and
\begin{eqnarray}
\label{eq:NeqExtWorkG}
\frac{d W_{\rm A}^{int} ( t )}{d t} \geq \frac{d G_{\rm A}^{\rm neq} ( t )}{d t} .
\end{eqnarray}
However, unlike in the Massieu--Planck case, the equality in Eqs. \eqref{eq:NeqExtWorkF} and \eqref{eq:NeqExtWorkG} cannot be obtained in general because the minimum DL work path in Fig. \ref{fig:Rel1} is evaluated by optimizing both $\beta(t)$ and $x(t)$, while $\beta$ has been fixed to introduce the Helmholtz energy as  $F^{\rm neq}_{\rm A} ( t ) = - \Phi_{\rm A}^{\rm neq} ( t ) / \beta$.

The above discussion indicates that the inclusion of the thermostatic process is a key to develop an efficient heat engine under non-equilibrium conditions.

\section{Convexity of non-equilibrium thermodynamic potentials as functions of extensive variables}
\label{sec:Convexity}

\begin{table*}[!t]
\caption{\label{table:Convexity} Inequalities between the non-equilibrium state at time $t_1$ and the equilibrium state at time $t_2$, with the conditions under which these inequalities hold. From these inequalities, we find that, for example, for a given inverse temperature and external field, the non-equilibrium Planck potential takes its maximum value when the state is in equilibrium.}

\begin{ruledtabular}
\begin{tabular}{lcc}
neq Potentials & Convexity and Concavity & Condition
\\
 \hline
Massieu & $(1 - \lambda ) ( \Phi_{\rm A}^{\rm neq} )_{A} + \lambda ( \Phi_{\rm A}^{\rm neq} )_{B}
\leq ( \Phi_{\rm A}^{\rm neq} )_\lambda$ & $\beta_{A} = \beta_{B} = \beta_{\lambda}$
\\
Planck & $( 1 - \lambda ) ( \Xi_{\rm A}^{\rm neq} )_{A} + \lambda ( \Xi_{\rm A}^{\rm neq} )_{B}
\leq ( \Xi_{\rm A}^{\rm neq} )_\lambda$ & $\beta_{A} = \beta_{B} = \beta_{\lambda}$ and $\tilde{x}_{A} = \tilde{x}_{B} = \tilde{x}_\lambda$
\\
Helmholtz & $( 1 - \lambda ) ( F_{\rm A}^{\rm neq} )_{A} + \lambda ( F_{\rm A}^{\rm neq} )_{B}
\geq ( F_{\rm A}^{\rm neq} )_\lambda$ & $\beta_{A} = \beta_{B} = \beta_{\lambda}$
\\
Gibbs & $( 1 - \lambda ) ( G_{\rm A}^{\rm neq} )_{A} + \lambda ( G_{\rm A}^{\rm neq} )_{B}
\geq ( G_{\rm A}^{\rm neq} )_\lambda$ & $\beta_{A} = \beta_{B} = \beta_{\lambda}$ and $x_{A} = x_{B} = x_\lambda$
\end{tabular}
\end{ruledtabular}

\end{table*}

\begin{figure}

\includegraphics[width=8cm]{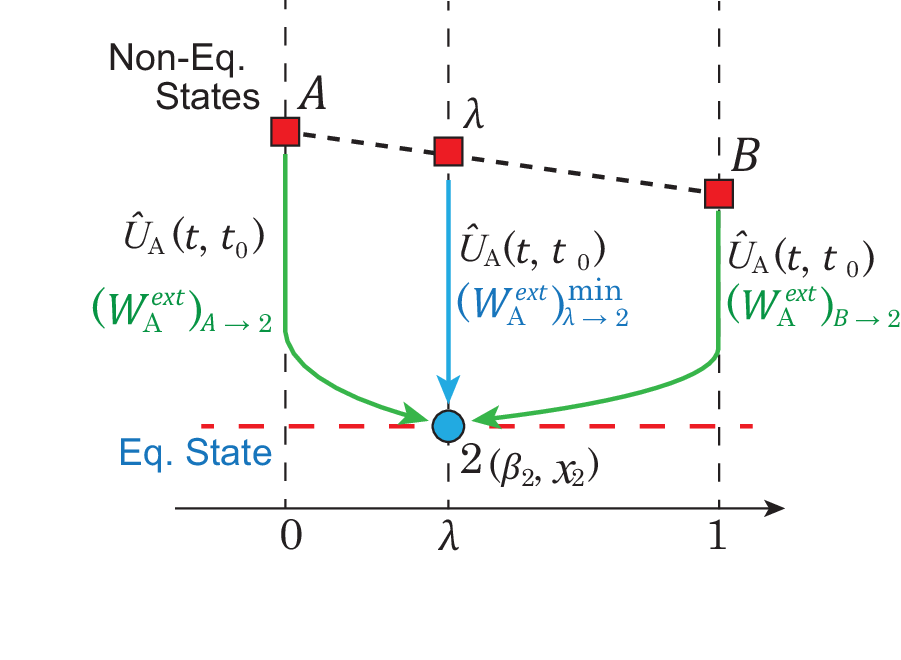}
\caption{\label{fig:Convex} Schematic for deriving convexity relations. The red squares represent the non-equilibrium states. The arrows are the relaxation paths to equilibrium state 2, as described by the time-evolution operator $\hat{U}_{\rm A} ( t , t_0 )$: The blue arrow is the DL minimum work path from state $\lambda$.}

\end{figure}

To obtain stable thermodynamic properties, the potential must be a convex function of the work variables. Even in the non-equilibrium case, we find that convexity holds under certain conditions. 
Here, we present this in the SB model. Because other potentials can be discussed similarly, we have limited our discussion to the case of non-equilibrium Massieu potential. 

We consider three non-equilibrium states $A$, $B$, and $\lambda$ ($\lambda \in [ 0 , 1 ]$) with the same initial inverse temperature $\beta ( t_0 )$  at time $t_0$. We then assume that the state $\lambda$ satisfies the  relation
\begin{eqnarray}
\label{eq:Convexity1}
(W_{\rm A})_\lambda = ( 1 - \lambda ) (W_{\rm A})_A + \lambda (W_{\rm A})_B ,
\end{eqnarray}
where $(W_{\rm A})_A=\left(W_{\rm A}(p, q; t) \right)_A$ and $ (W_{\rm A})_B=\left(W_{\rm A}(p, q; t) \right)_B$ are the Wigner distributions in the quantum case (or phase space distributions in the  classical case) in the $A$ and $B$ states. The time evolution operator of a subsystem is expressed as $\hat{U}_{\rm A} ( t , t_0 ) = \exp\!\big[ - \int \hat{\mathscr L} ( p, q )\,dt \big]$, where $\hat{\mathscr L} ( p, q )$ denotes the Liouville operator, which includes fluctuation and dissipation operators for equilibration. In the quantum case, $\hat{\mathscr L} ( p, q )$ acts on $W_{\rm A}$, which is represented by hierarchical elements.\cite{T06JPSJ, T20JCP, IT19JCTC}
We then obtain the relations between the extensive variable and internal energies as
\begin{eqnarray}
\label{eq:Convexity2}
({X}_{\mathrm{A}})_{\lambda} ( t ) = ( 1 - \lambda )({X}_{\mathrm{A}})_{A}(t) 
+ \lambda ({X}_{\mathrm{A}})_{B}  ( t ) 
\end{eqnarray}
and
\begin{eqnarray}
\label{eq:Convexity3}
(U_{\mathrm{A}})_{\lambda} ( t ) = ( 1 - \lambda ) (U_{\mathrm{A}})_{A}( t ) + \lambda (U_{\mathrm{A}})_{B} ( t ) ,
\end{eqnarray}
where we have defined $( X_{\mathrm{A}} )_\alpha ( t ) = \mathrm{tr}_{\rm tot} \{ X_{\rm A} ( q ) \hat{U}_{\rm tot} ( t , t_0 ) ( W_{\rm tot} )_\alpha \}$ and $( U_{\mathrm{A}} )_\alpha ( t ) = \mathrm{tr}_{\rm tot} \{ [ p^2 / 2 m + U ( q ) ] \hat{U}_{\rm tot} ( t , t_0 ) ( W_{\rm tot} )_\alpha \}$ for $\alpha = A , B,$ and $\lambda$. Thus, from Eq.~\eqref{eq:DefWext0}, the DL extensive work $( W_{\rm A}^{ext} )_\alpha ( t , t_0 )$ ($\alpha = A , B,$ and $ \lambda$), performed between  times $t_0$ and $t$ also satisfies the equality
\begin{eqnarray}
\label{eq:Convexity4}
( W_{\rm A}^{ext} )_{\lambda} ( t , t_0 ) = ( 1 - \lambda ) ( W_{\rm A}^{ext} )_A ( t , t_0 ) \nonumber \\
+ \lambda ( W_{\rm A}^{ext} )_B ( t , t_0 ). 
\end{eqnarray}

The DL minimum work path from the state $\lambda$ to the equilibrium state 2 is described by $\hat{U}_{\rm A} ( t , t_0 )$ (see Fig.~\ref{fig:Convex}). This indicates that when $t$ is large, the intensive variables must be $\beta ( t ) \rightarrow \beta_2$ and ${x} ( t ) \rightarrow x_2$, where $\beta_2$ and $x_2$ are the values at the equilibrium state 2.
Therefore, processes starting from $A$ and $B$ also relax to the equilibrium state 2 as $t \rightarrow \infty$. Then, from Eq.~\eqref{eq:Convexity4}, we obtain
\begin{eqnarray}
\label{eq:Convexity5}
( W_{\rm A}^{ext} )_{\lambda \rightarrow 2}^{\rm min} = ( 1 - \lambda ) ( W_{\rm A}^{ext} )_{A \rightarrow 2}
+ \lambda ( W_{\rm A}^{ext} )_{B \rightarrow 2} .
\end{eqnarray}
Using the inequality $( W_{\rm A}^{ext} )_{\alpha \rightarrow 2} \geq ( W_{\rm A}^{ext} )_{\alpha \rightarrow 2}^{\rm min}$ for the states $\alpha = A$ and $B$ and Eq.~\eqref{eq:A2}, we obtain the inequality to prove convexity as
\begin{eqnarray}
\label{eq:Convexity6}
( 1 - \lambda ) ( \Phi_{\rm A}^{\rm neq} )_A + \lambda ( \Phi_{\rm A}^{\rm neq} )_B
\leq ( \Phi_{\rm A}^{\rm neq} )_\lambda ,
\end{eqnarray}
where $( \Phi_{\rm A}^{\rm neq} )_\alpha$ is the non-equilibrium Massieu potential in the state $\alpha$ ($\alpha = A$, $B$, and $\lambda$).
When $A$ and $B$ are in equilibrium, we obtain the convexity relation in the quasi-static case for a fixed inverse temperature using the inequality for the Massieu potential as Table~\ref{table:InequalityQstNeqDL} as \cite{KT24JCP1}
\begin{eqnarray}
\label{eq:QstConvex}
( 1 - \lambda ) ( \Phi_{\rm A}^{\rm qst} )_A + \lambda ( \Phi_{\rm A}^{\rm qst} )_B
\leq ( \Phi_{\rm A}^{\rm qst} )_\lambda .
\end{eqnarray}

The convexity and concavity properties for the other non-equilibrium thermodynamic potentials are summarized in Table~\ref{table:Convexity}.

\section{Inequality between  non-equilibrium and quasi-static thermodynamic potentials}
\label{sec:neq-qstIneq}


\begin{table*}[!t]

\caption{\label{table:InequalityQstNeqDL} Inequalities for  Massieu--Planck potentials between the non-equilibrium state $A$ at time $t_1$ and the equilibrium state 2 at time $t_2$,  together with the conditions under which these inequalities hold. From these inequalities, we find that, for example, for a given inverse temperature and intensive variable, the non-equilibrium Planck potential takes its maximum value when $A$ is in the equilibrium state.}

\begin{ruledtabular}
\begin{tabular}{lcc}
neq Potentials & Inequalities & Condition
\\
 \hline
Massieu & $\Phi_{\rm A}^{\rm qst} ( t_2 ) \geq \Phi_{\rm A}^{\rm neq} ( t_1 )$
& $\beta ( t_1 ) = \beta ( t_2 )$ and $X_{\rm A} ( t_1 ) = X_{\rm A} ( t_2 )$
\\
Planck  & $\quad \Xi^{\rm qst}_{\rm A} ( t_2 ) \geq \Xi_{\rm A}^{\rm neq} ( t_1 ) \quad$
& $\beta ( t_1 ) = \beta ( t_2 )$ and $\tilde{x} ( t_1 ) = \tilde{x} ( t_2 )$
\\
C--Entropy & $\Lambda_{\rm A}^{\rm qst} ( t_2 ) \geq \Lambda_{\rm A}^{\rm neq} ( t_1 )$
& $U_{\rm A} ( t_1 ) = U_{\rm A} ( t_2 )$ and $X_{\rm A} ( t_1 ) = X_{\rm A} ( t_2 )$
\\
B--Entropy & $\Gamma_{\rm A}^{\rm qst} ( t_2 ) \geq \Gamma_{\rm A}^{\rm neq} ( t_1 )$
& $U_{\rm A} ( t_1 ) = U_{\rm A} ( t_2 )$ and $\tilde{x} ( t_1 ) = \tilde{x} ( t_2 )$
\end{tabular}
\end{ruledtabular}
\end{table*}

\begin{table*}[!t]

\caption{\label{table:InequalityQstNeq}Inequalities for  Helmholtz--Gibbs potentials between the non-equilibrium state at time $t_1$ and the equilibrium state at time $t_2$, with the conditions under which these inequalities hold.
}

\begin{ruledtabular}
\begin{tabular}{lcc}
neq Potentials & Inequalities & Condition
\\
 \hline
Helmholtz & $F_{\rm A}^{\rm qst} ( t_2 ) \leq F_{\rm A}^{\rm neq} ( t_1 )$
& $T ( t_1 ) = T ( t_2 )$ and $X_{\rm A} ( t_1 ) = X_{\rm A} ( t_2 )$
\\
Gibbs & $G_{\rm A}^{\rm qst} ( t_2 ) \leq G_{\rm A}^{\rm neq} ( t_1 )$
& $T ( t_1 ) = T ( t_2 )$ and $x ( t_1 ) = x ( t_2 )$
\\
 Internal Energy  & $U_{\rm A} ( t_2 ) \leq U_{\rm A} ( t_1 )$
&  $S_{\rm A}^{\rm neq} ( t_1 ) = S_{\rm A}^{\rm qst} ( t_2 )$ and $X_{\rm A} ( t_1 ) = X_{\rm A} ( t_2 )$ 
\\
Enthalpy & $H_{\rm A} ( t_2 ) \leq H_{\rm A} ( t_1 )$
& $S_{\rm A}^{\rm neq} ( t_1 ) = S_{\rm A}^{\rm qst} ( t_2 )$ and $x ( t_1 ) = x ( t_2 )$
\end{tabular}
\end{ruledtabular}

\end{table*}


Under given conditions, the non-equilibrium DL thermodynamic potentials are smaller than the quasi-static ones. Here, we derive the inequalities between the quasi-static and non-equilibrium thermodynamic potentials. 

We consider a non-equilibrium-to-equilibrium transition $A\rightarrow 2$ from time $t_1$ to $t_2$ described by the inverse temperature $\beta ( t)$ and the extensive variable $X_{\rm A}^{\rm neq} (t)$. 
The DL extensive work in this process is evaluated as
\begin{eqnarray}
\label{eq:AppD1}
 \int^{t_2}_{t_1} \left[ U_{\rm A} ( t' ) \frac{d \beta ( t' )}{d t'} - \tilde{x} ( t' ) \frac{d X_{\rm A}^{\rm neq} ( t' )}{d t'} \right] d t' \nonumber \\
 = \tilde{x}_2 ( X_{\rm A}^{\rm neq} ( t_2 ) - X_{\rm A}^{\rm neq} ( t_1 ) ), 
\end{eqnarray}
where we have assumed that $\beta ( t ) = \beta_2$ for time $t \ge t_1$ and $\tilde{x} ( t ) = \tilde{x}_2$ for time $t > t_1$ with the the intensive variables $\beta_2$ and $\tilde{x}_2$, allowing the subsystem to relax to the equilibrium state 2 at time $t_2$. The value of $\tilde{x}_2$ is set to satisfy the condition $X_{\rm A}^{\rm neq} ( t_2 ) = X_{\rm A}^{\rm neq} ( t_1 )$.
Because  $X_{\rm A}^{\rm neq} ( t_2 ) = ( X_{\rm A}^{\rm qst} )_2 = X_{\rm A}^{\rm neq} ( t_1 )$, the right-hand side of  Eq.~\eqref{eq:AppD1} vanishes. Thus, the net DL extensive work is zero, and so  the DL minimum work principle reduces to $0 \geq ( \Phi_{\rm A}^{\rm neq} )_{A \rightarrow 2}$, from which it follows that $\Phi_{\rm A}^{\rm qst} ( t_2 ) \geq \Phi_{\rm A}^{\rm neq} ( t_1 )$. 

The inequalities for the other DL thermodynamic potentials were obtained in the same manner. The results are summarized in Tables~\ref{table:InequalityQstNeqDL} and~\ref{table:InequalityQstNeq}

\begin{figure}

\includegraphics[width=8cm]{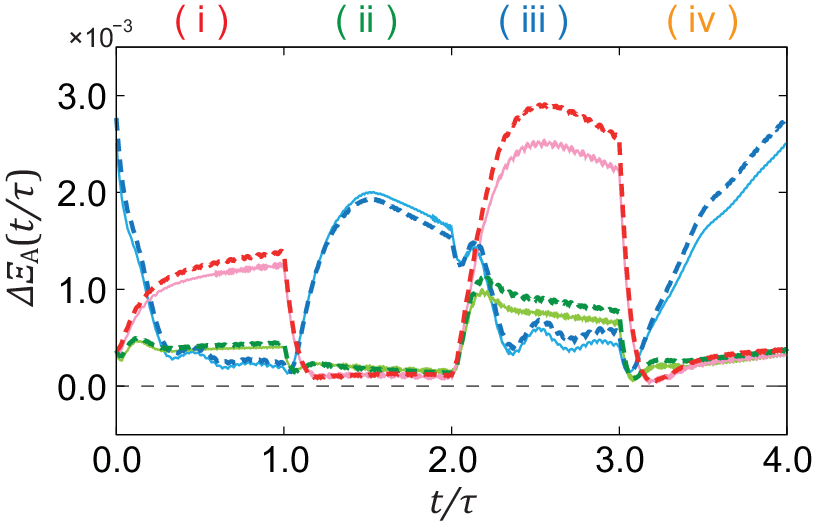}
\caption{\label{fig:Planck}Difference between the quasi-static and non-equilibrium Planck potentials in the quantum and classical cases (solid and dashed curves, respectively) for one cycle of the thermostatic Stirling engine. The blue, green, and red curves represent the weak ($A = 0.5$), intermediate ($A = 1.0$), and strong ($A = 1.5$) SB coupling, respectively.}

\end{figure}

In Fig.~\ref{fig:Planck}, to illustrate the inequalities presented in Table~\ref{table:InequalityQstNeqDL}, we plot the difference between the quasi-static and non-equilibrium Planck potential, $\Delta \Xi_{\rm A} ( t ) = \Xi_{\rm A}^{\rm qst} [ \beta ( t ) , \tilde{x} ( t ) ] - \Xi_{\rm A}^{\rm neq} ( t )$, for  one cycle of the thermostatic Stirling engine described in Sec.~\ref{sec:StirlingEngine}.  As can be seen, $\Delta \Xi_{\rm A} ( t )$ is always positive and satisfies the inequalities in Table~\ref{table:InequalityQstNeqDL}.

\bibliography{references,tanimura_publist}
\end{document}